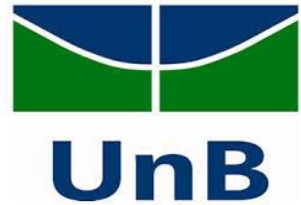

# Additional Tests for TV 3.0

______________________________

August 2024

University of Brasília (UnB, *Universidade de Brasília*)

# Table of Contents

# 1 Introduction

In 2023 we have conducted extensive experiments on subjective video quality for the TV 3.0 project at University of Brasília [1]. A full report on these tests is available at the Fórum SBTVC website (https://forumsbtvd.org.br/wp-content/uploads/2024/03/SBTVD-TV_3_0-P3-VC-Report.pdf). These tests have evaluated the H.266/VVC codec and a hybrid codec formed by the H.266/VVC [2] and the LCEVC (Low Complexity Enhancement Video Coding) [3] with different resolutions, ranging from 720p to 4K. Table 1 shows the use cases (Video Under Test – VUTs) tested in these experiments, as well as the reference video and the bitrates tested.

*Table 1 - Configurations used for the 2023 tests*

| Test | Reference Video | Video Under Test (VUT) | Bitrates |
|---|---|---|---|
| VUT 1.1 | 1 080i H.264/AVC + SL-HDR1 | 720p H.266/VVC HDR10 | [1.50 2.24 3.35 5.02 7.50] Mbps |
| VUT 1.2 | 1 080i H.264/AVC + SL-HDR1 | 1 080p H.266/VVC HDR10 | [2.00 2.99 4.47 6.69 10.00] Mbps |
| VUT 1.3 | 1 080p H.266/VVC HDR10 | 2 160p H.266/VVC HDR10 | [4.00 5.98 8.94 13.37 20.00] Mbps |
| VUT 2.3 | 1 080i H.264/AVC + SL-HDR1 | 1 080p H.266/VVC HDR10 (Base Layer) | [2.40 3.60 5.40 8.00 12.00] Mbps |
| VUT 2.4 | 1 080p H.266/VVC HDR10 | 2 160p [LCEVC (Enhancement Layer) + 1 080p H.266/VVC HDR10 (Base Layer)] | [2.67 3.99 5.96 8.92 13.33] Mbps |

In some of the meetings of the SBTVD forum two questions were raised about these experiments, when comparing the two VUTs in 4K resolution (2160p), namely VUT 1.3 and VUT 2.4. The first question was that the underlying VVC codecs had different implementations – VUT 1.3 tested a VVC implementation from Ateme, known as TitanLive, while VUT 2.4 tested a VVC implementation from MainConcept, known as MainConcept encoder. The second question was that VUT 1.3 and VUT 2.4 had different coding configurations. VUT 1.3 used 1 intra frame every two seconds, for a GOP size of 120 frames, while VUT 2.4 used 1 intra frame every 6 seconds, for a GOP size of 360 frames. Therefore, it was difficult to pinpoint if the results reported in the 2023 experiments were due to a difference in GOP size, a difference in the VVC implementation, or a difference between the codecs themselves.

For these reasons, we have been contacted by V-Nova to run additional experiments. This new experiment consists of two new VUTs, one with the VVC codec at 4K resolution, and the other with the VVC+LCEVC codec at 4K resolution. Naturally, both codecs have the same GOP size (120 frames) and use the same VVC encoder (MainConcept live encoder). This new experiment follows the same experimental protocol as the previous experiments, in order to be fully comparable to the reported results. It is important to note that, apart from providing the encoders, V-Nova has not interfered with the experiments in any way. This document details the results of the new experiments.

# 2 Research Team

This report was developed by the following team at the University of Brasília:

    Eduardo Peixoto Fernandes da Silva, Universidade de Brasília - http://lattes.cnpq.br/7720922538480924

    Pedro Garcia Freitas, Universidade de Brasília - http://lattes.cnpq.br/0669244312752630

    Mylene Christine Queiroz Farias, Universidade de Brasília - http://lattes.cnpq.br/4465619366143200

    José Edil Guimarães de Medeiros, Universidade de Brasília - http://lattes.cnpq.br/2307490672260330

    Gabriel Correia Lima da Cunha e Menezes, Universidade de Brasília - http://lattes.cnpq.br/0543326076269968

    André Henrique Macedo da Costa, Universidade de Brasília - http://lattes.cnpq.br/4223927174237834

We also had the collaboration of SBTVD Forum members, in particular:

    Carlos Cosme – Chair, Audio & Video Coding Workgroup, SBTVD Forum

# 3 Glossary

| | |
|---|---|
| 1 080i | Interlaced video with a resolution of 1 920 x 1 080 pixels |
| 1 080p | Progressive video with a resolution of 1 920 x 1 080 pixels |
| 2 160p | Progressive video with a resolution of 3 840 x 2 160 pixels |
| 4K | Progressive video with a resolution of 3 840 x 2 160 pixels |
| AVC | Advanced Video Coding |
| BT.2020 exchange | Recommendation ITU-R BT.2020: Parameter values for ultra-high-definition television systems for production and international programme |
| CRF | Constant Rate Factor |
| fps | frames per second |
| GOP | Group of Pictures |
| H.264 | Recommendation ITU-T H.264: Advanced video coding for generic audiovisual services |
| H.266 | Recommendation ITU-T H.266: Versatile video coding |
| HDR | High Dynamic Range |
| HDR10 | High Dynamic Range 10-bit |
| ITU-R | International Telecommunications Union - Radiocommunication sector |
| ITU-T | International Telecommunications Union - Telecommunication standardization sector |
| kbps | kilobits per second |
| LCEVC | Low Complexity Enhancement Video Coding |
| MaxCLL | Maximum Content Light Level |
| MaxFALL | Maximum Frame Average Light Level |
| Mbps | Megabits per second |
| MOS | Mean Opinion Score |
| MPEG | Motion Picture Experts Group |
| OLED | Organic Light-Emitting Diode |
| SBTVD | *Sistema Brasileiro de Televisão Digital* (Brazilian Digital Television System) |
| SDI | Serial Digital Interface |
| SI | Spatial Index |
| SL-HDR | Single Layer HDR |

| SMPTE | Society of Motion Picture and Television Engineers |
|---|---|
| TI | Temporal Index |
| TV | TeleVision |
| VQEG | Video Quality Experts Group |
| VUT | Video Under Test |
| VVC | Versatile Video Coding |

# 4 Real-Time Video Coding Subjective Quality Assessment Results

The experiment protocol is the same as the tests reported to the forum in the TV 3.0 Phase 3 (https://forumsbtvd.org.br/wp-content/uploads/2024/03/SBTVD-TV_3_0-P3-VC-Report.pdf). All encoders, both for the reference video and for the Video Under Test, were used in a real-time configuration, and the input video was fed to the encoder using a SDI connection. Only the overall quality of the video streams is evaluated, and the participants are not asked why they have rated the videos in a certain way.

## 4.1 Experimental Setup

This section details the contents used, the laboratory conditions, and the interface of the experiments.

### 4.1.1 Experimental Stimuli

The video dataset used for the experiments was prepared by the SBTVD Forum specifically for the tests. It comprises five different video streams, mostly camera-generated content with some computer-generated graphical overlays and high contrast colorful scenes, each one with a few scene cuts. Some content contains smooth gradients between similar colors to assist in detecting possible banding artifacts. Some of the content have scenes commonly transmitted through Brazilian television, such as football matches and carnival scenes. The scenes were chosen due to their diverse range of spatial and temporal information. The details of the video streams are shown in Table 2.

The spatial and temporal information for the scenes is shown in Figure 1, while Figure 2 shows the file size for each content when encoding with libx265 at CRF = 28. This is used as an indicator of the complexity of encoding each content. It can be seen that content globo01 needs almost four times the bitrate compared to content philips03 when using the same Quantization Parameter. A short description and a sample frame for each content are shown in Table 3.

*Table 2 - Details of the contents used for the tests*

| Resolution (horizontal x vertical) | 3 840 x 2 160 |
|---|---|
| Aspect ratio | 16:9 (square pixels) |
| Sampling ratio | $YC_BC_R$ 4:2:2 |
| Bit depth | 10 bits |
| Frame rate | 59.94 fps |
| Scan | Progressive |
| Transfer characteristics | Perceptual Quantization (max 1 000 cd/m²) |
| HDR Metadata | HDR10 (SMPTE ST 2086, MaxFALL and MaxCLL) |
| Color Gamut | Rec. ITU-R BT.2020 |
| Duration | 30 s |
| File format | Uncompressed MOV file |

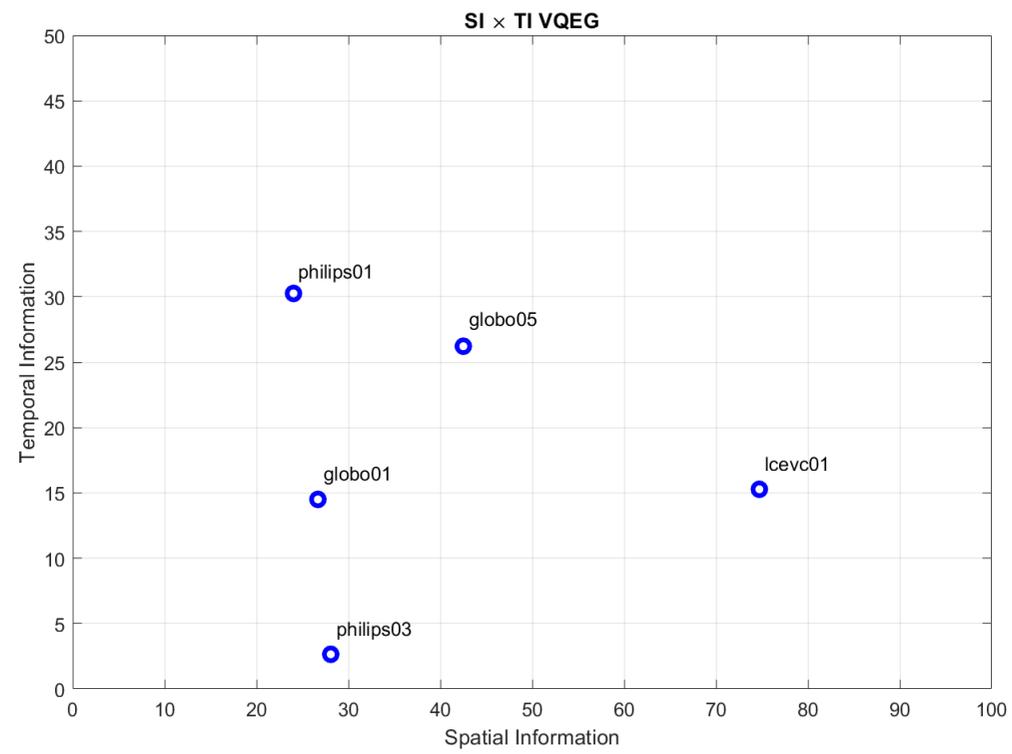

*Figure 1 - Spatial and Temporal Information for the test contents used*

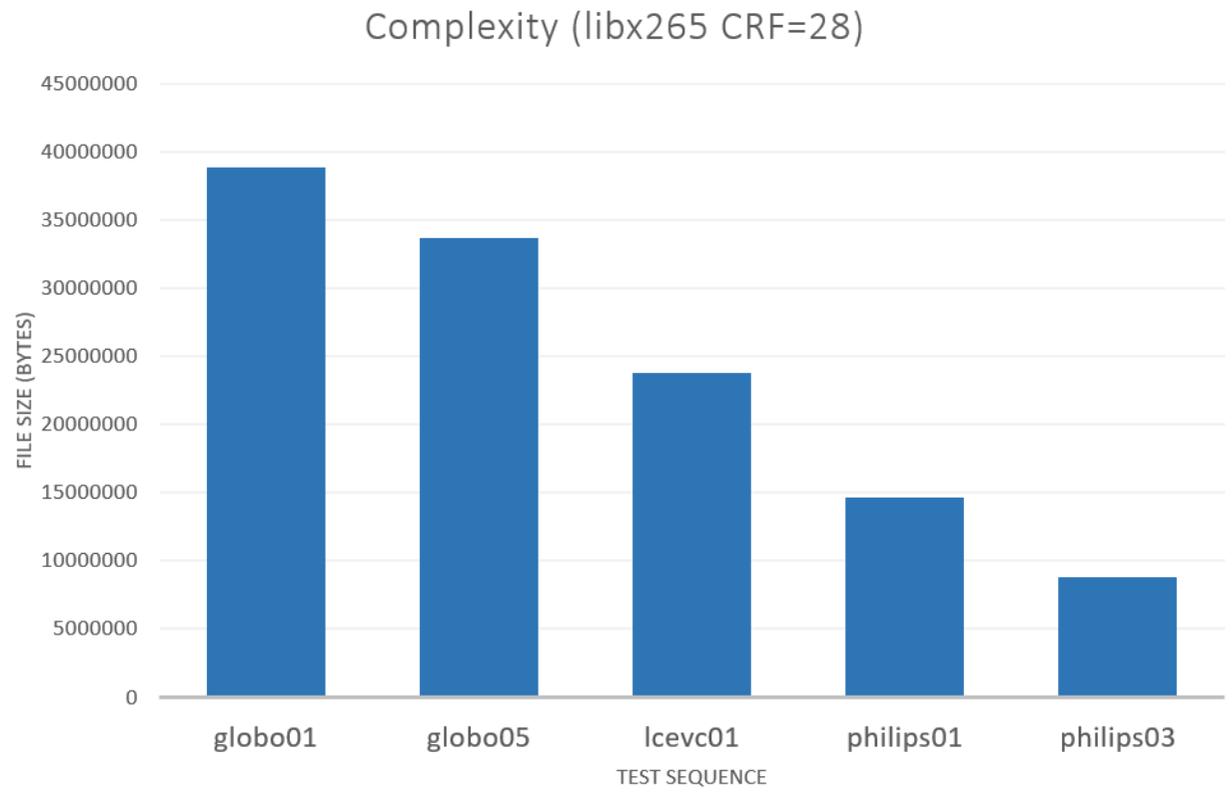

*Figure 2 - Content file size when encoded with libx265 at CRF=28*

*Table 3 - Short description of the contents used for the tests*

| | | |
|---|---|---|
| **globo01** | 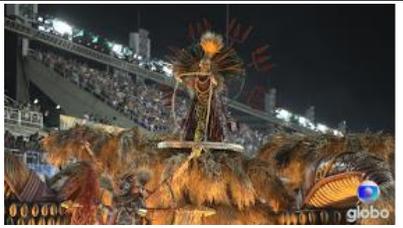 | Excerpts from the 2022 Rio de Janeiro samba school parade. This is a high-movement colorful scene. There is a lot of content with high spatial information, including many human faces close to the camera. |
| **globo05** | 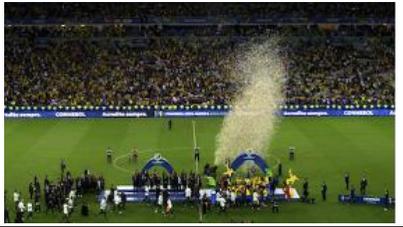 | Excerpts from the 2019 Conmebol Copa America (South America Football Championship) final match. There is a lot of high-speed movement and a large number of people moving around in the background. |
| **lcevc01** | 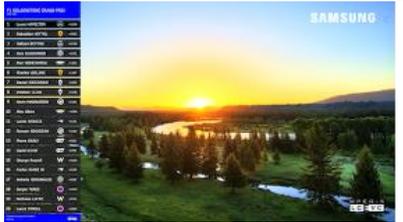 | A collection of short clips of different landscapes with direct sunlight and graphics overlay, including a few sunrise and sunset scenes, and a sunflower field with bees. There are also some Formula 1 results graphics overlay in the left part of the screen which contributes to a lot of high-frequency content. |

| philips01 | 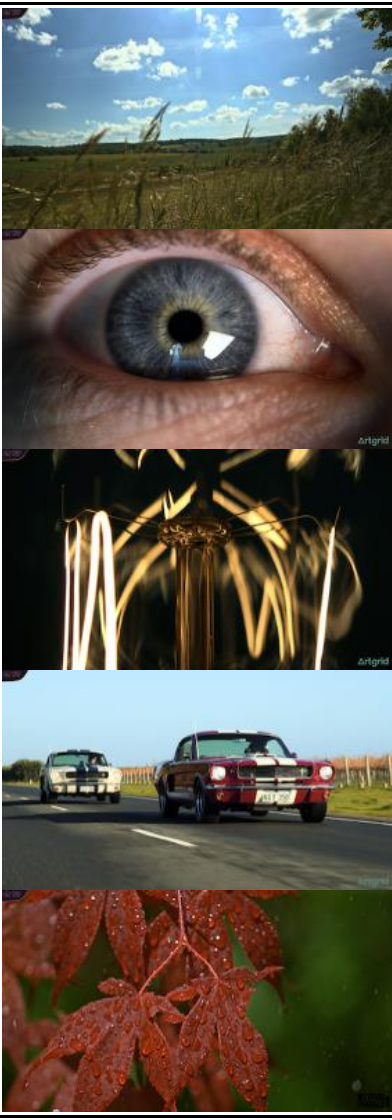 | A collection of short clips, including a wheat field, a close-up of a human eye, a close-up of an incandescent lamp, fast-moving vintage cars on a highway, and some bright red leaves with raindrops. |

| | |
|---|---|
| **philips03** 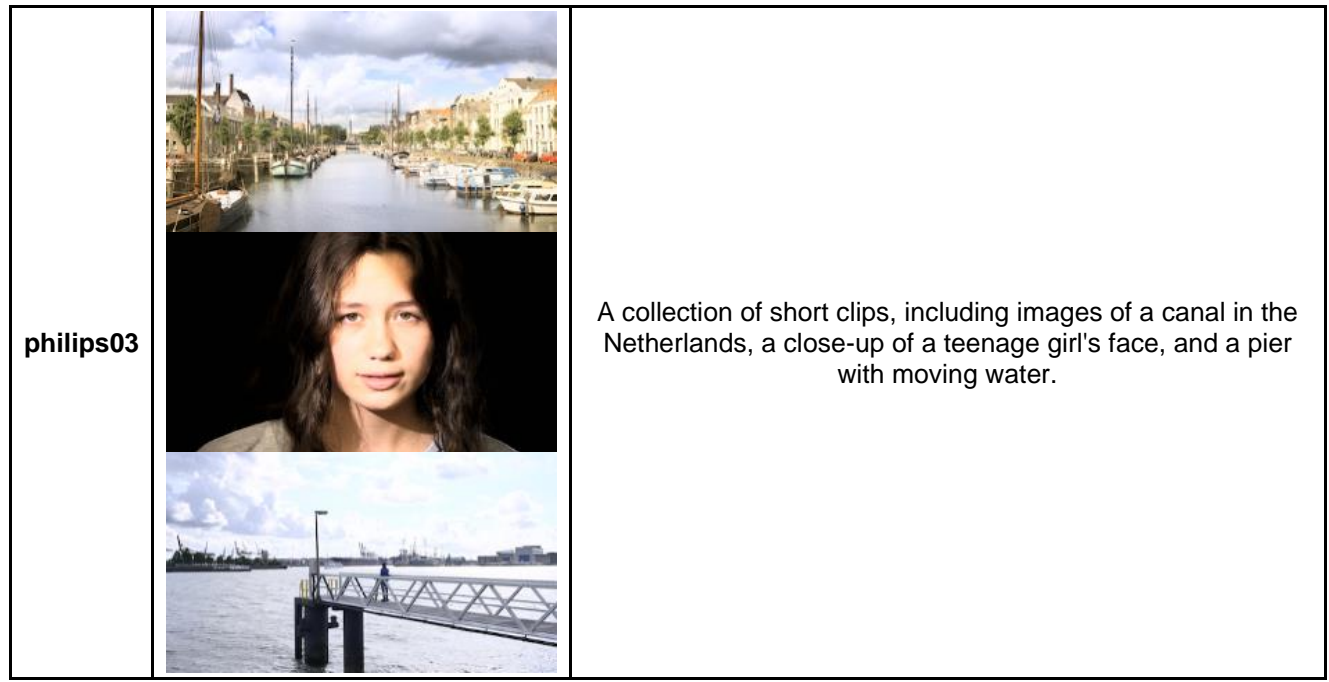 | A collection of short clips, including images of a canal in the Netherlands, a close-up of a teenage girl's face, and a pier with moving water. |

### 4.1.2 Experimental Conditions

The experiments were performed on a 465cm x 490cm laboratory located on the first floor of the SG-11 building at the University of Brasília Darcy Ribeiro Campus. The lab was available exclusively for the TV 3.0 Phase 3 experiments.

Figure 3 shows the relative position of the subject in relation to the screen. In this figure, H is the vertical size of the screen, which is 81 cm. Therefore, the viewing distance from the screen was around 131 cm.

The equipment setup included two sets of the following equipment:

1. Mac Studio (2022 model) desktop equipped with an Apple M1 Ultra and 128GB of RAM running MacOS Ventura 13.2.1 operating system and a Blackmagic Design Media Express 3.8.1 software;
2. Blackmagic Design UltraStudio 4k Mini; and
3. LG 65" OLED TV set model OLED65C2PSA.CWZQLJZ (firmware version 03.33.85) with a screen area of approximately 81cm x 142cm.

The LG TVs are consumer-grade products; thus, a limited set of configurations was available. Table 4 shows the configurations used for both TV sets.

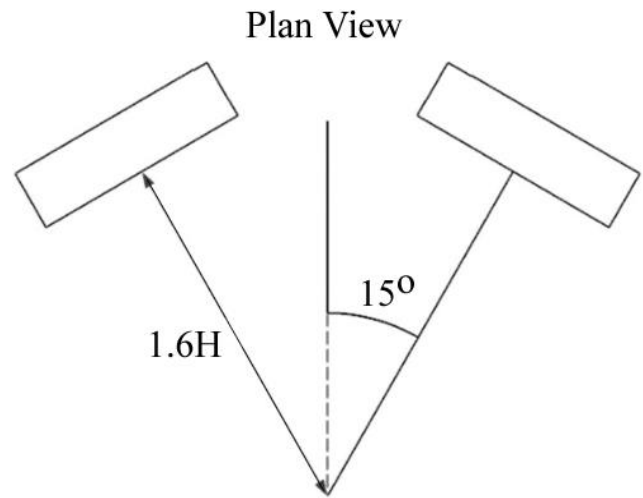

*Figure 3 - Laboratory set-up for the subjective tests*

*Table 4 - TV sets configuration*

| Parameter | | Setting |
|---|---|---|
| Brightness | OLED Pixel Brightness | 80 |
| | Adjust Contrast | 95 |
| | Black Level | 50 |
| | Auto Dynamic Contrast | Medium |
| | Dynamic Tone Mapping | On |
| | Peak Brightness | High |
| | Gamma | 2.2 |
| | Video Range | Auto |
| | Motion Eye Care | Off |
| Color | Color Depth | 55 |
| | Tint | 0 |
| | Color Gamut | Auto Detect |
| Color/Fine Tune | Color Adjustment | Off |
| Balance | Color Temperature | 0 |
| | Method | 2 Points |
| | Point | High |
| | Red | 0 |
| | Green | 0 |
| | Blue | 0 |
| Clarity | Adjust Sharpness | 25 |
| | Super Resolution | Off |
| | Noise Reduction | Off |
| | MPEG Noise Reduction | Off |
| | Smooth Gradation | Off |
| | Real Cinema | Off |
| | TruMotion | Off |

It should be noted that the TV sets used in the test have a progressive 3 840 x 2 160 display. The TV set upscales any input signal with a lower resolution to its display native resolution.

Table 5 shows the configuration used for the 2023 tests, and Table 6 shows the configuration used for the new 2024 tests. The new tests were named as VUT 1.4 and VUT 2.5 for consistency.

*Table 5 - VUTs used in the 2023 tests*

| Test | Reference Video | Video Under Test (VUT) | Bitrates |
|---|---|---|---|
| VUT 1.1 | 1 080i H.264/AVC + SL-HDR1 | 720p H.266/VVC HDR10 | [1.50 2.24 3.35 5.02 7.50] Mbps |
| VUT 1.2 | 1 080i H.264/AVC + SL-HDR1 | 1 080p H.266/VVC HDR10 | [2.00 2.99 4.47 6.69 10.00] Mbps |
| VUT 1.3 | 1 080p H.266/VVC HDR10 | 2 160p H.266/VVC HDR10 | [4.00 5.98 8.94 13.37 20.00] Mbps |
| VUT 2.3 | 1 080i H.264/AVC + SL-HDR1 | 1 080p H.266/VVC HDR10 (Base Layer) | [2.40 3.60 5.40 8.00 12.00] Mbps |
| VUT 2.4 | 1 080p H.266/VVC HDR10 | 2 160p [LCEVC (Enhancement Layer) + 1 080p H.266/VVC HDR10 (Base Layer)] | [2.67 3.99 5.96 8.92 13.33] Mbps |

*Table 6 - VUTs used in the new 2024 tests*

| Test | Reference Video | Video Under Test (VUT) | Bitrates |
|---|---|---|---|
| VUT 1.4 | 1 080p H.266/VVC HDR10 | 2 160p H.266/VVC HDR10 | [4.00 5.98 8.94 13.37 20.00] Mbps |
| VUT 2.5 | 1 080p H.266/VVC HDR10 | 2 160p [LCEVC (Enhancement Layer) + 1 080p H.266/VVC HDR10 (Base Layer)] | [2.67 3.99 5.96 8.92 13.33] Mbps |

### 4.1.3 Experimental Protocol

We have performed 2 different independent tests, each one with at least 30 subjects, properly screened for normal visual acuity and normal color vision. For each test, we have used 5 different contents, provided by the SBTVD Forum and detailed in Section 4.1.1. For each test, 5 different bitrates were used for the video under test (VUT), while the reference video was kept constant for all bitrates. The goal of each test was to find the bitrate at which a particular quality was achieved for all contents.

For all tests, the reference was always displayed on the left screen, and this was known by the participant. For each VUT, we have used three different pseudo-random sequences to display the 25 videos of each session.

The experimental session comprises two distinct segments: training and subjective evaluation. In the initial training phase, the conductor outlines the experiment's objectives and procedures, gathers demographic information from participants, and conducts a training session to get them acquainted with the experimental tasks. Before beginning the experiment, subjects are asked to read a 4-page document and fill in some information. The first two pages of the document contain an explanation of the experimental protocol and the risks associated with it. Subsequently, participants are requested to sign a consent form and provide details about their age, gender, technical expertise, and familiarity with visual impairments.

In the training session, each subject is shown 3 video samples with varying quality levels that correspond to different encoding parameters. The goal of the training session is to provide the subjects an opportunity to familiarize themselves with the experimental task and to understand the quality rating scale. After the training, the experimenter may answer questions from the subject. Then, the experimenter takes note of the VUT information as well as the pseudo-random sequence number (used to analyze the data) and starts the subjective evaluation phase.

During the subjective evaluation phase, participants assess a series of 25 test video sequences, each lasting around 30 seconds. To record their quality scores, participants utilize paper forms and pens. The evaluation form, based on the stimulus-comparison method with adjectival categorical judgment as described in Recommendation ITU-R BT.500-15 [4][5], Section A4-4.1 from Annex 4 to Part 2, contains a 7-level quality scale, as depicted in Table 7, for each test sequence to be evaluated. Prior to the presentation of each video, a 3-second preparation countdown is displayed on a gray screen. Following each video, a gray screen prompts participants to complete the evaluation form. The pace of the video exhibition is manually controlled by the experimenter, who adjusts the experiment's speed based on the capacities of the participants.

*Table 7 - Subjective evaluation scale*

| Textual Scale (in Portuguese) | Textual Scale | Numeric Scale |
|---|---|---|
| Muito Pior | Much Worse | -3 |
| Pior | Worse | -2 |
| Pouco Pior | Slightly Worse | -1 |
| Igual | Equal | 0 |
| Pouco Melhor | Slightly Better | 1 |
| Melhor | Better | 2 |
| Muito Melhor | Much Better | 3 |

## 4.2 Data Analysis Protocol

This section details how the experimental data gathered during the experiments was processed. In the first part, the subjective scores for each VUT were processed to generate an output target bitrate. This target bitrate is the minimum bitrate for which a target quality score is achieved for all contents. Then, the videos are encoded at this target bitrate and the output bitstream is analyzed in order to assess the minimum and maximum bitrates to be used by the codec configuration. To achieve these target bitrates, some processing steps are performed. These steps include removing outliers and computing the mean opinion scores (MOS), performing an experimental model regression, and analyzing the video stream data.

### 4.2.1 Outlier Removal and Mean Opinion Score Determination

In conducting the statistical analysis of the experimental data for this study, we used the following protocol to ensure the reliability and validity of the obtained results. The protocol involves several key steps, starting with the identification and removal of outliers. Preliminary data inspection used statistical measures, such as the computation of Z-scores, to flag potential outliers. In this work, outliers are defined as data points lying outside predetermined thresholds and are an indicator of measurement errors. The outlier detection step was based on Recommendation ITU-R BT.500-15. Following the recommendation, the threshold used was 2 standard deviations. The subsequent removal of these outliers was performed systematically to prevent their undue influence on the subsequent statistical analysis phases. Following outlier removal, a critical step of the statistical analysis is the transformation of the collected experimental scores. Then, the results were further analyzed by finding the mean opinion scores (MOS) and confidence intervals (CI), as described in Recommendation ITU-T P.1401.

### 4.2.2 Experimental Model Regression

After removing the outliers and determining the MOS for each video, a regression to a logistic function is applied to enhance the interpretability and reliability of the scores. This transformation adjusts raw scores, improving the model fit to the underlying experimental data. The ultimate goal of this transformation is to provide an accurate representation of the relationship between objective and subjective variables, capturing non-linear patterns. In this context, we fitted a logistic function to the MOS versus bitrate relationship. The logistic function is a monotonic function given by the following equation:

$$MOS_p = \frac{b_1}{1 + exp(-b_2 \times (bitrate - b_3))},$$

where $MOS_p$ is the predicted MOS as a function of the *bitrate* and $b_1$, $b_2$, and $b_3$ are parameters obtained via least squares fit[1]. In summary, the above predictive model was used to evaluate various metrics on the sets of subjectively measured MOS and the corresponding bitrate. Using this model, we predicted the bitrates where a given encoded video is better than a reference video, comparing the studied codecs.

---

[1] https://www.itu.int/md/T01-SG09-C-0060

### 4.2.3   Video Stream Analysis Protocol

Once a target bitrate is found for each VUT, each content was encoded at the defined target bitrate using the same codec and parameters as that VUT. We analyze the bitstream to find the minimum and maximum bitrates of each GOP. This is not the ideal methodology to analyze the video stream bitrate variation over time and should be considered only as an indication of this variation.

## 4.3  VUTs

The details for each VUT are given in this section. The VUTs for the new experiment are named VUT 1.4 and VUT 2.5 as they share similarities with VUTs 1.3 and 2.4 of the 2023 experiments, respectively.

### 4.3.1   VUT 1.4 Definition

The goal of this VUT was to test the VVC encoder working at 2 160p resolution. The reference video for this VUT was encoded with the VVC at 1 080p resolution, at 7.52 Mbps. Note that the reference video used is the same reference video used for the VUTs 1.3 and 2.4 in the 2023 experiment. The complete details are found in Table 8. Since the VUT was encoded with a higher resolution than the reference video, the quality target considered was "the same" (0). Note that using the highest bitrate corresponding to the target score 0 ("the same") among the five clips in the test material means that this VUT would provide a similar subjective quality for the clip with the highest required bitrate, and somewhat higher score in the other clips while not necessarily reaching the score 1 ("slightly better") in all clips.

*Table 8 - VUT 1.4 encoding details*

|  | **Reference Video** | **VUT 1.3** |
|---|---|---|
| **Content label** | 1 080p | 2 160p |
| **Resolution** | 1 920 x 1 080 | 3 840 x 2 160 |
| **Frame rate** | 59.94 fps | 59.94 fps |
| **Scan** | Progressive | Progressive |
| **Bit depth** | 10 bits | 10 bits |
| **Color gamut** | BT.2020 | BT.2020 |
| **HDR Mode** | HDR10 | HDR10 |
| **Codec** | H.266/VVC | H.266/VVC |
| **GOP size** | 120 frames (2 seconds) | 120 frames (2 seconds) |
| **Encoder** | Ateme TitanLive Innovation v 4.1.31.911 | MainConcept Live Encoder v 0.0.0.4036 |
| **Encoder type** | Real-time | Real-time |
| **Bitrate** | 7.52 Mbps | [4.00 5.98 8.94 13.37 20.00] Mbps |

### 4.3.1.1 VUT 1.4 Experimental Findings

**Threshold σ=-1**

*Table 9 - VUT 1.4 results targeting a MOS grade of -1 (slightly worse)*

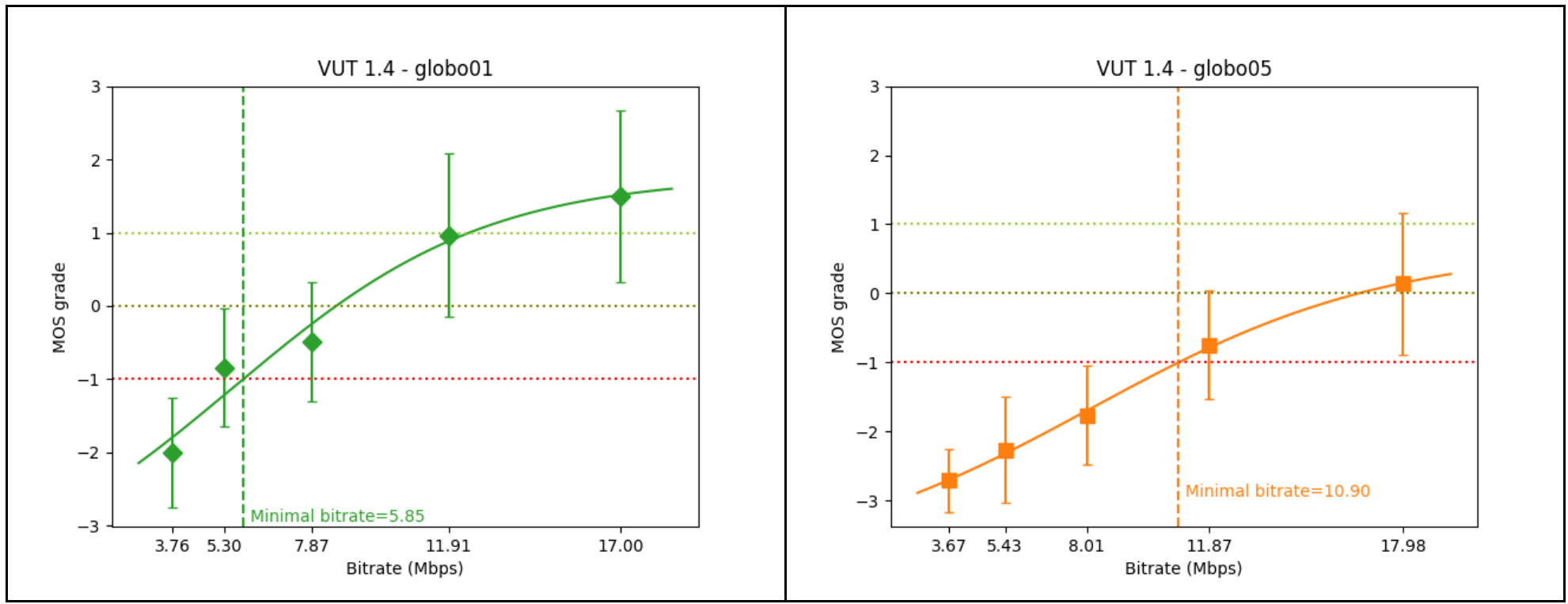

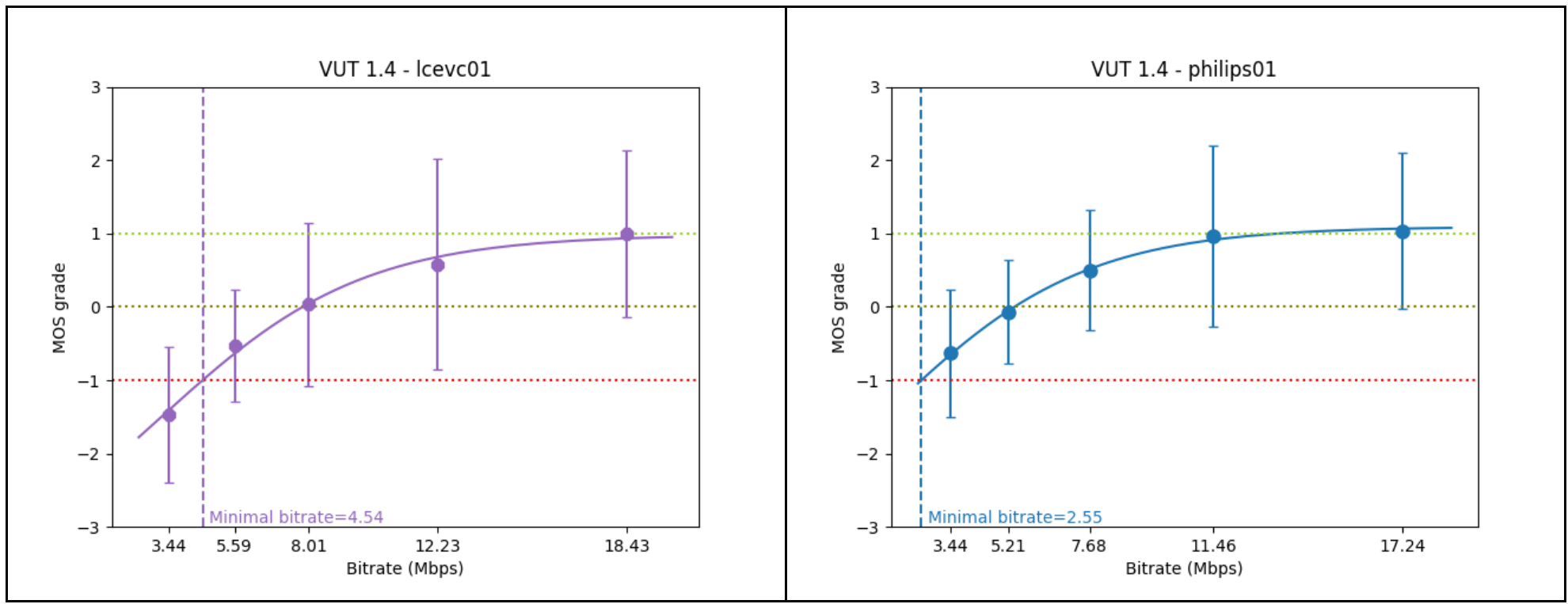

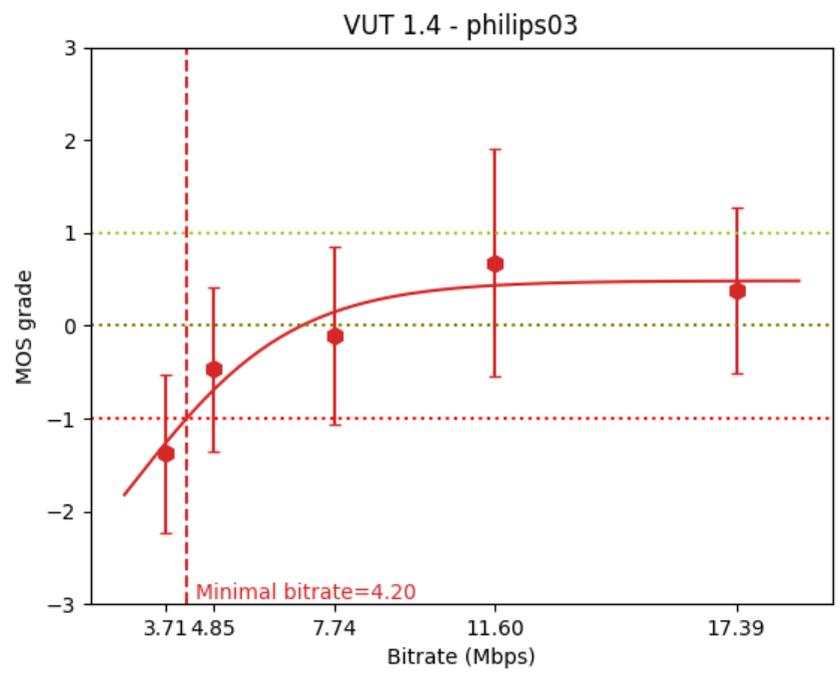

**Threshold σ=0**

Table 10 - VUT 1.4 results targeting a MOS grade of 0 (same quality)

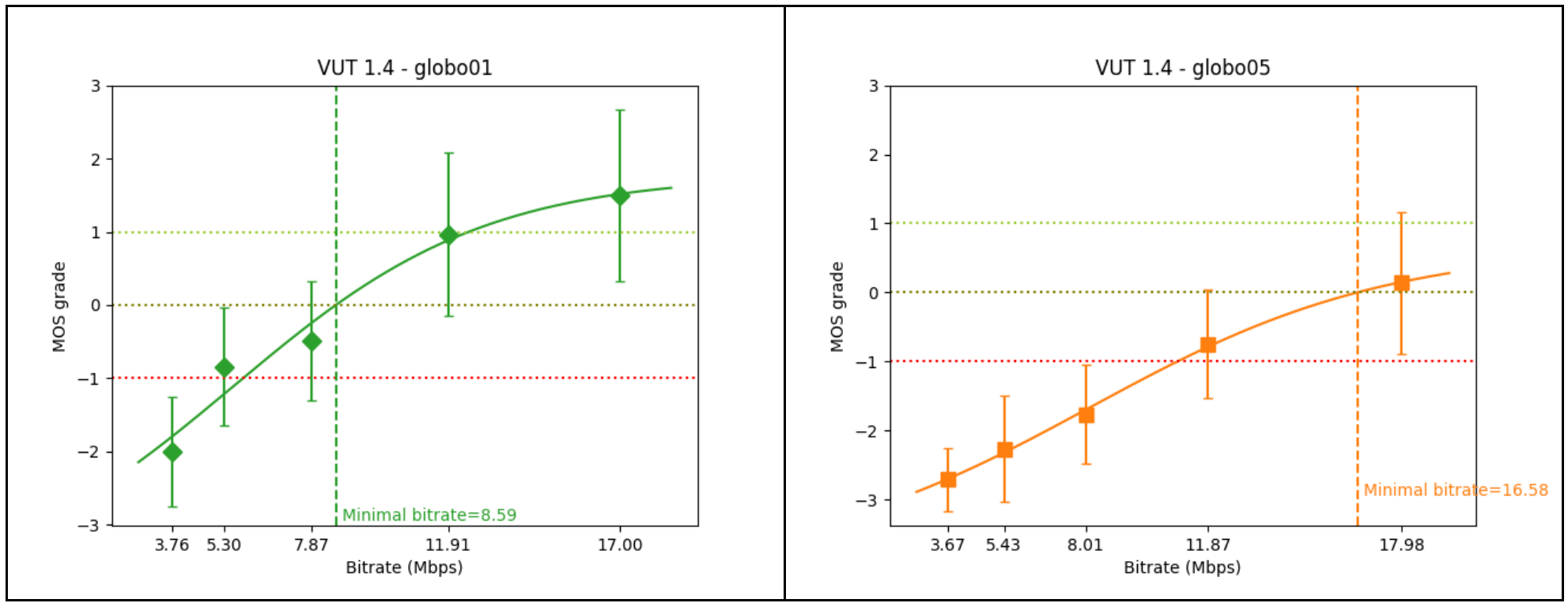

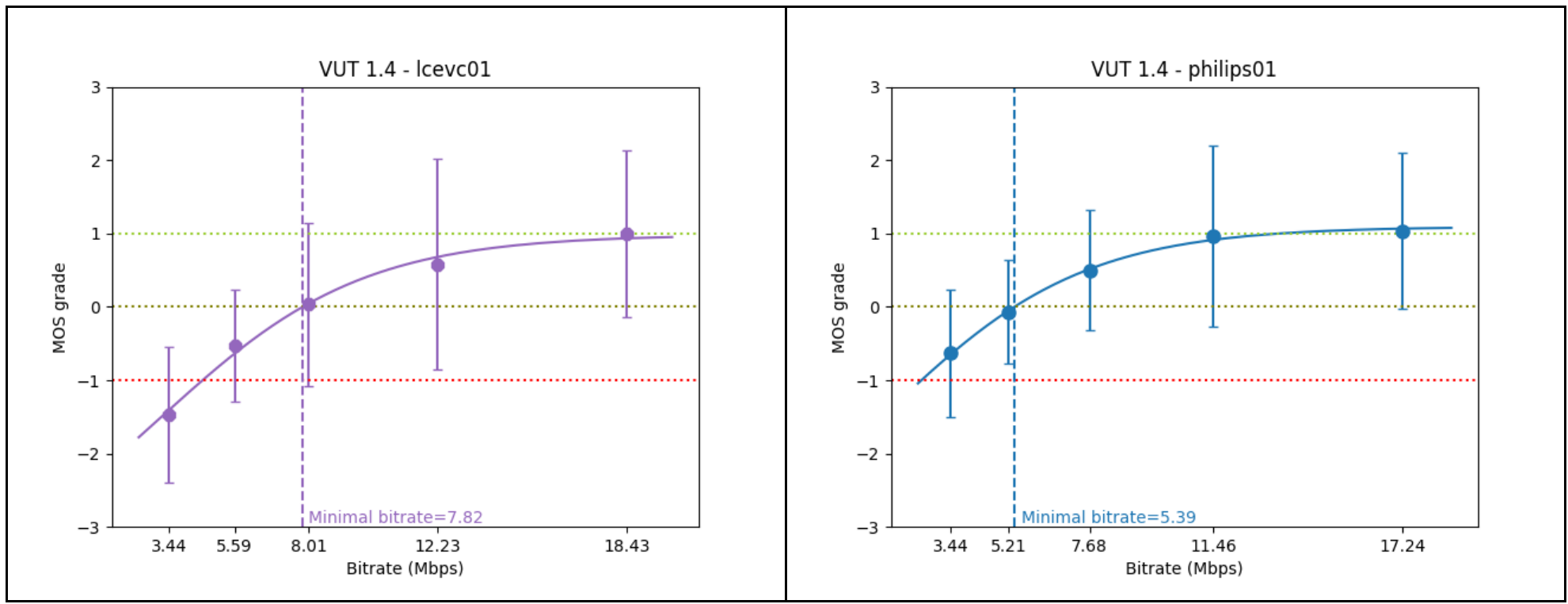

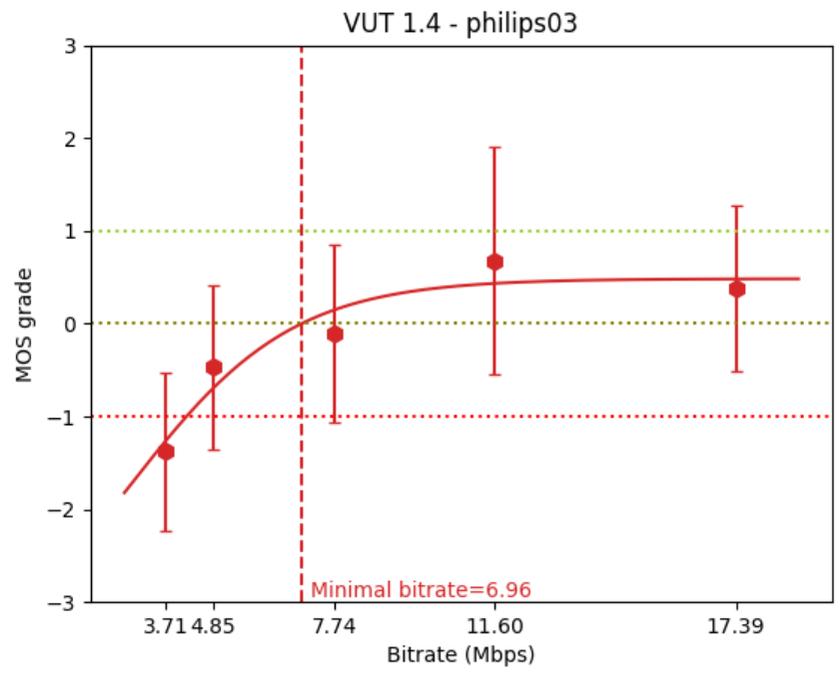

### Threshold σ=1

*Table 11 - VUT 1.4 results targeting a MOS grade of 1 (slightly better)*

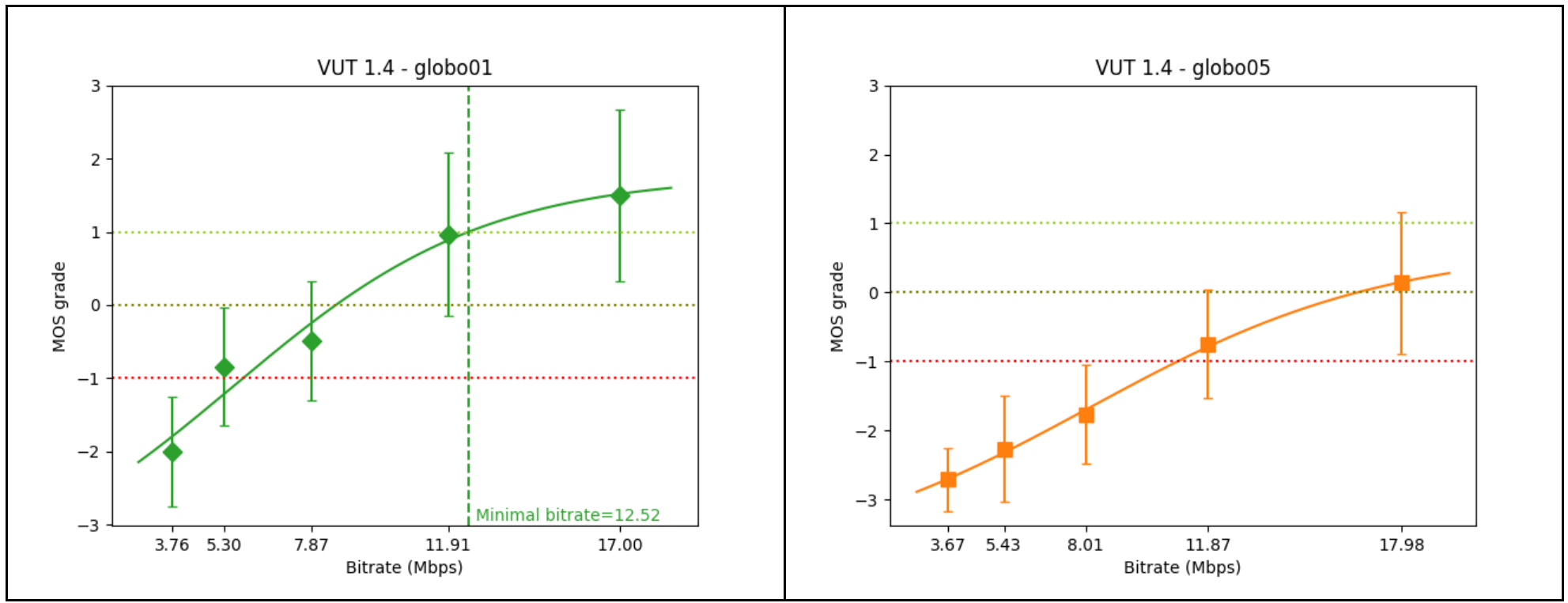

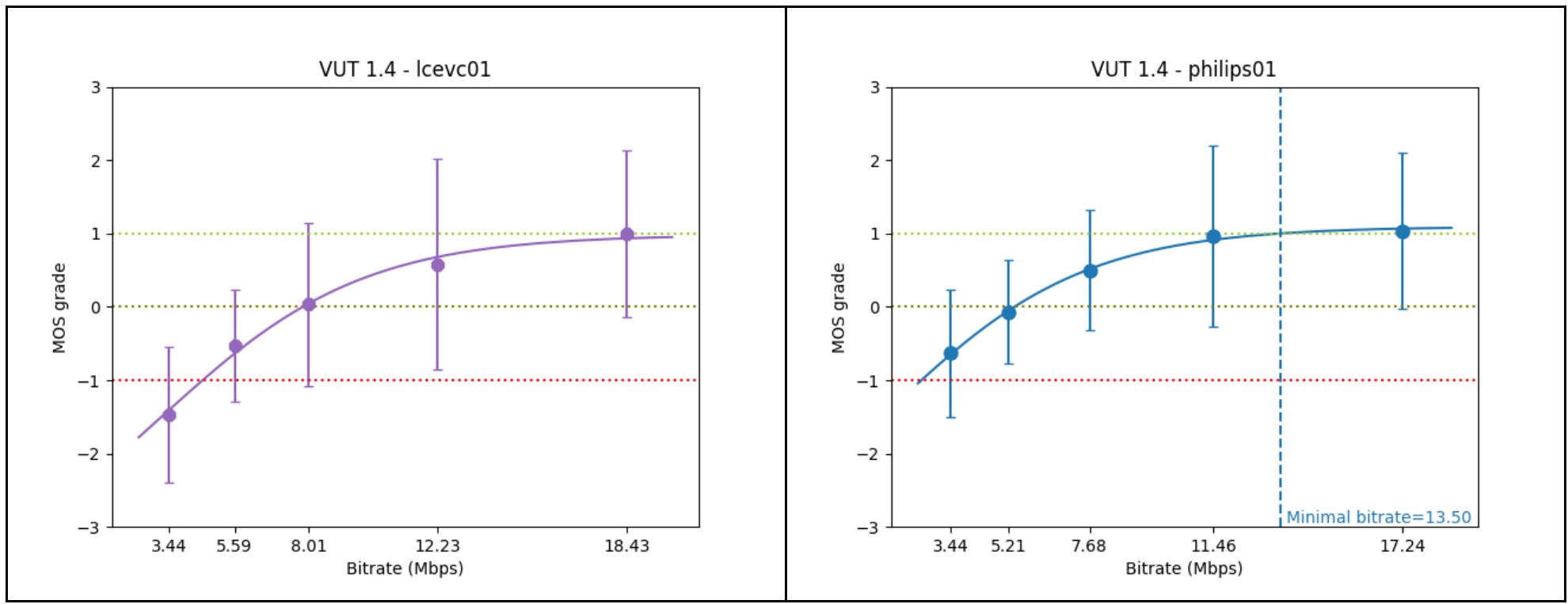

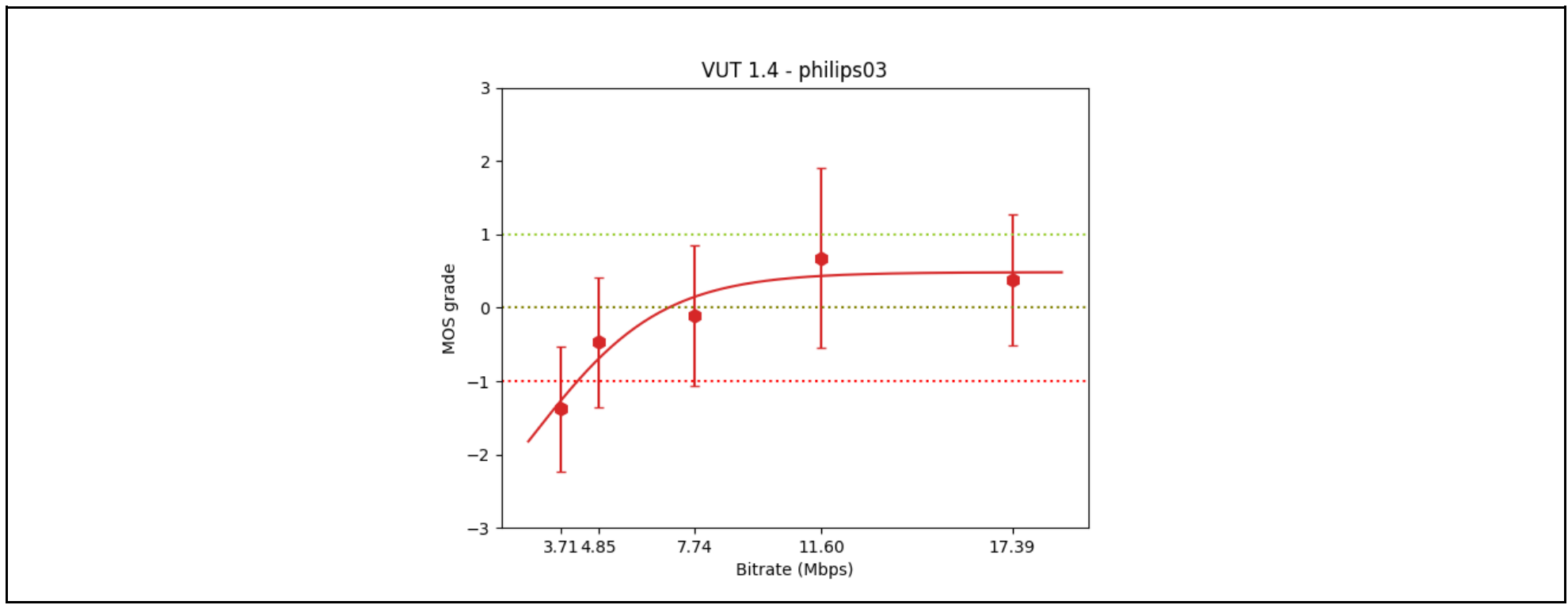

### 4.3.1.2 VUT 1.4 Output bitrate analysis

The bitrates for which each target quality is achieved are shown in Table 12.

Table 12 - Achieved Target Bitrates per content (the output target bitrate is highlighted)

| Target Quality | globo01 | globo05 | Icevc01 | philips01 | philips03 |
|---|---|---|---|---|---|
| -1 | 5.85 Mbps | **10.90 Mbps** | 4.54 Mbps | 2.55 Mbps | 4.20 Mbps |
| 0 | 8.59 Mbps | **16.58 Mbps** | 7.82 Mbps | 5.39 Mbps | 6.96 Mbps |
| 1 | 12.52 Mbps | Not achieved | Not achieved | Not achieved | Not achieved |

Following the results from the previous sections, the output bitrate for VUT 1.4 was found to be 16.58 Mbps.

### 4.3.1.3 VUT 1.4 Bitrate per GOP analysis

We have also carried out an analysis of the bitrate per GOP for each of the bitstreams used in the tests. This is a simple analysis, where the bitrate used for each GOP (i.e., between two intra frames) is computed. It must be noted that this analysis is a naïve way to test the bitrate output by an encoder.

**Bitrate per GOP analysis for target bitrate of 4000 kbps**

*Table 13 - VUT 1.4 Bitrate per GOP when encoded with the output bitrate of 4.00 Mbps*

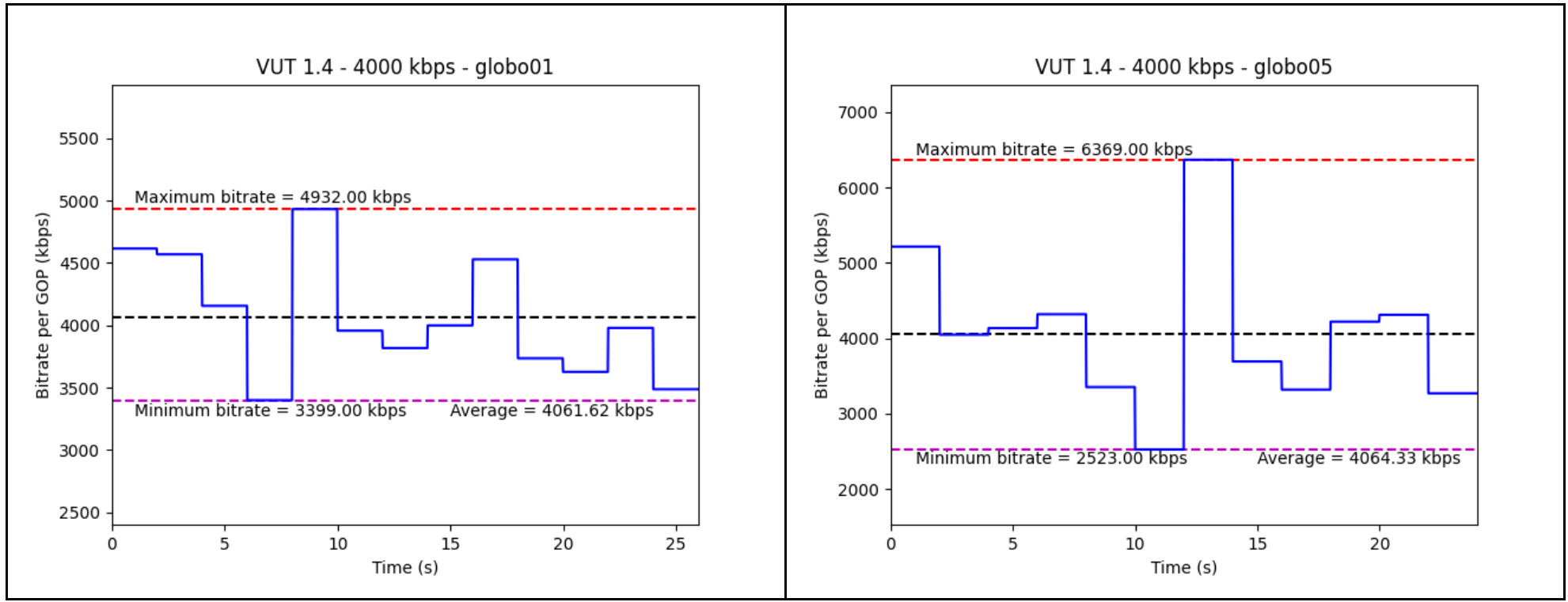

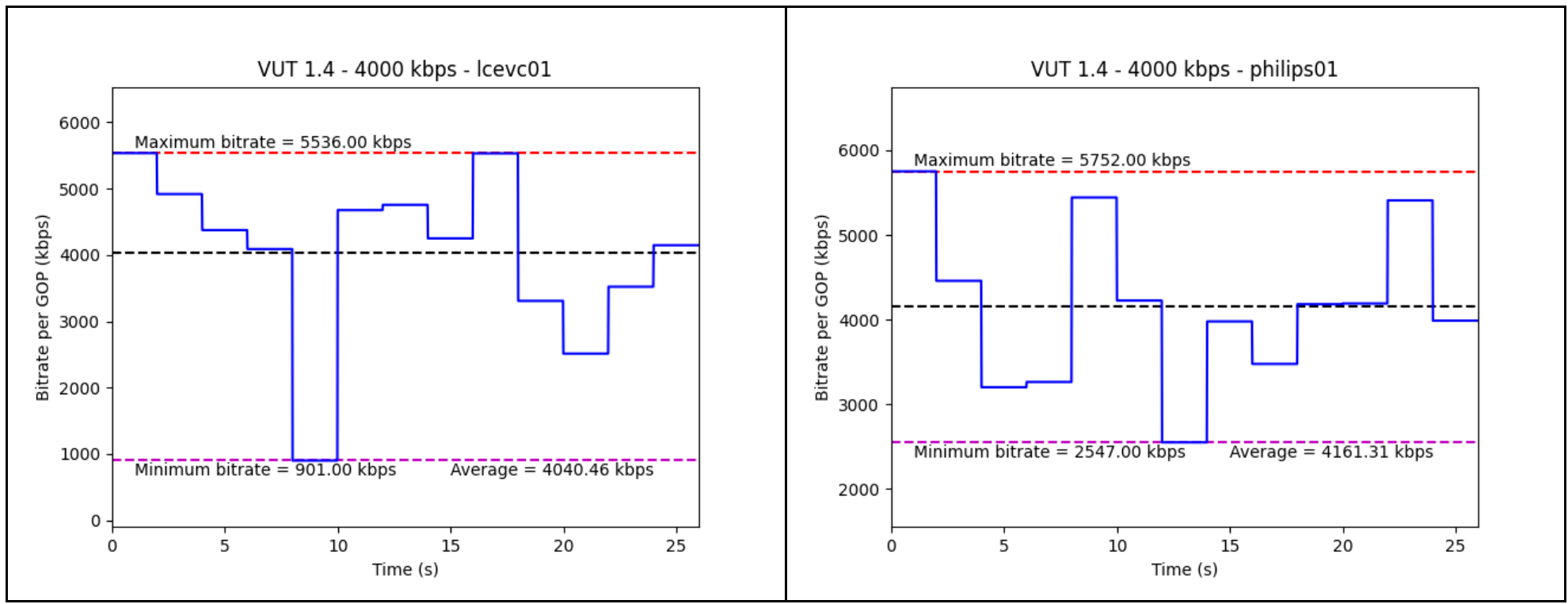

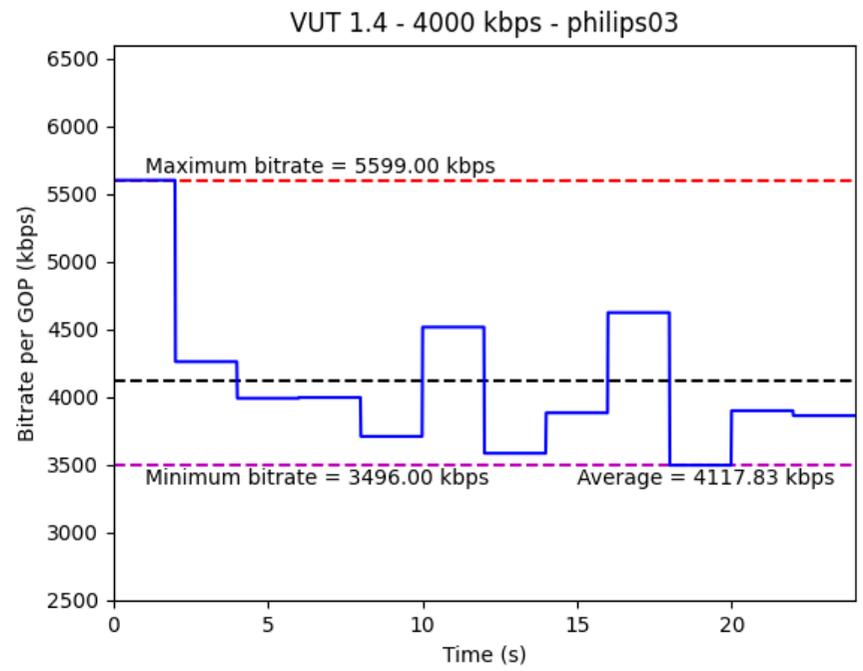

**Bitrate per GOP analysis for target bitrate of 5980 kbps**

*Table 14 - VUT 1.4 Bitrate per GOP when encoded with the output bitrate of 5.98 Mbps*

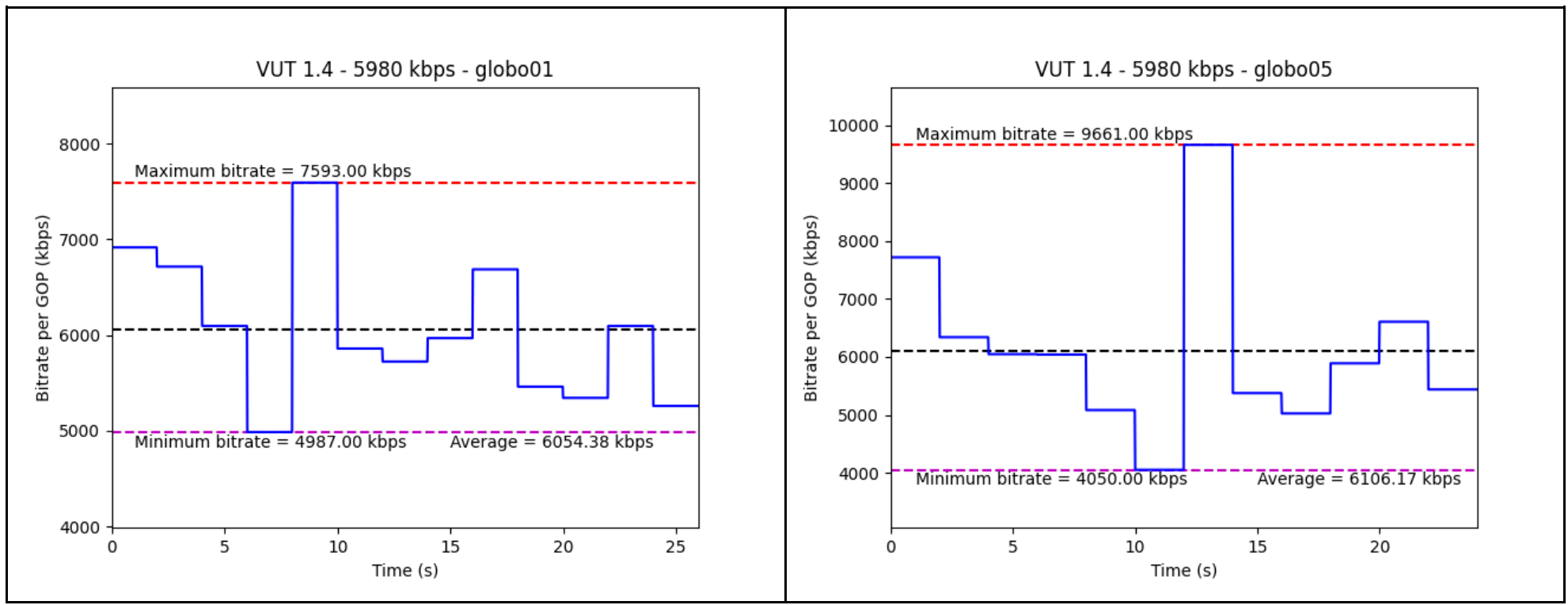

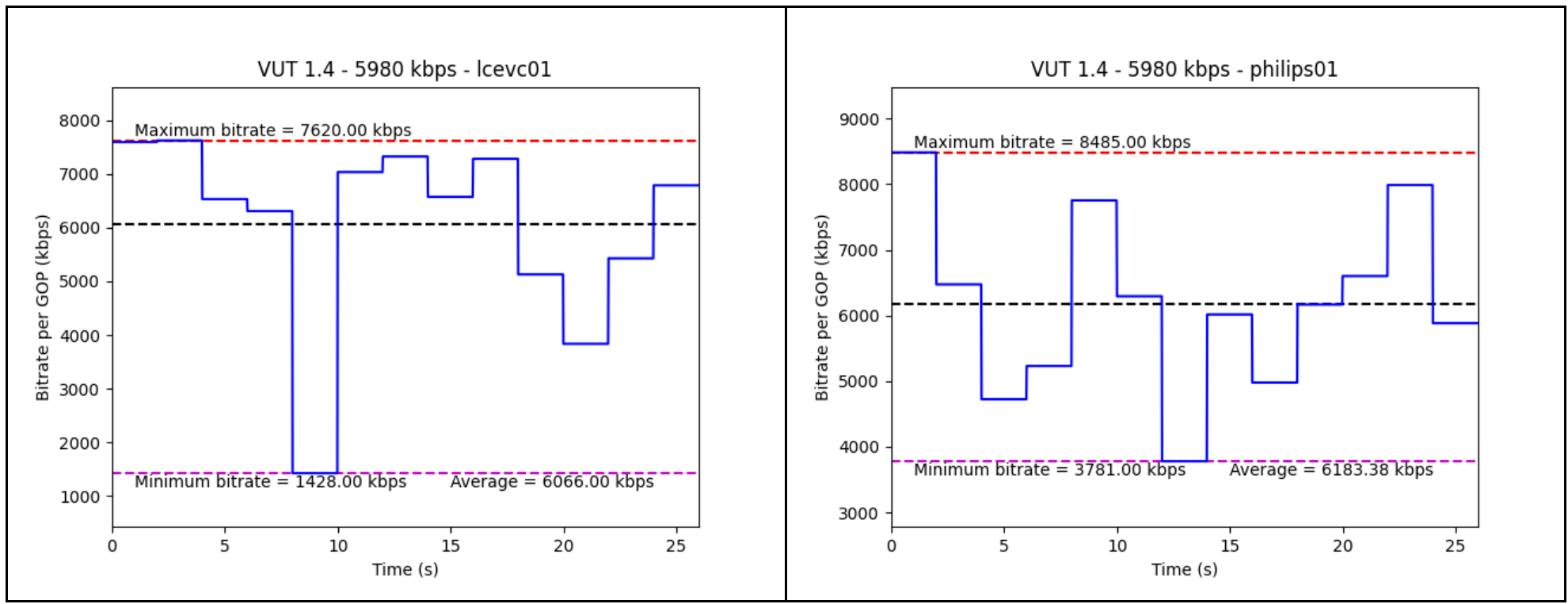

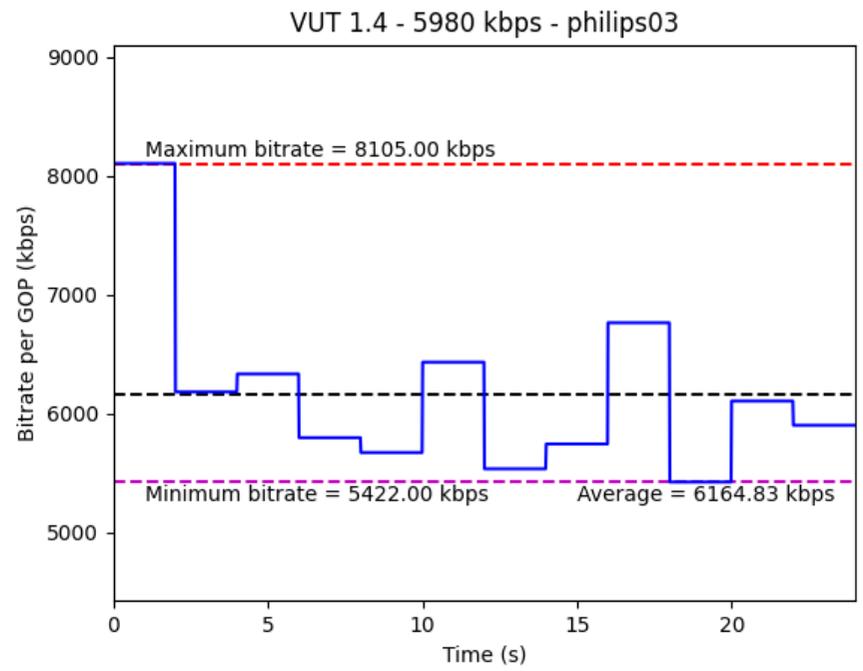

**Bitrate per GOP analysis for target bitrate of 8940 kbps**

*Table 15 - VUT 1.4 Bitrate per GOP when encoded with the output bitrate of 8.94 Mbps*

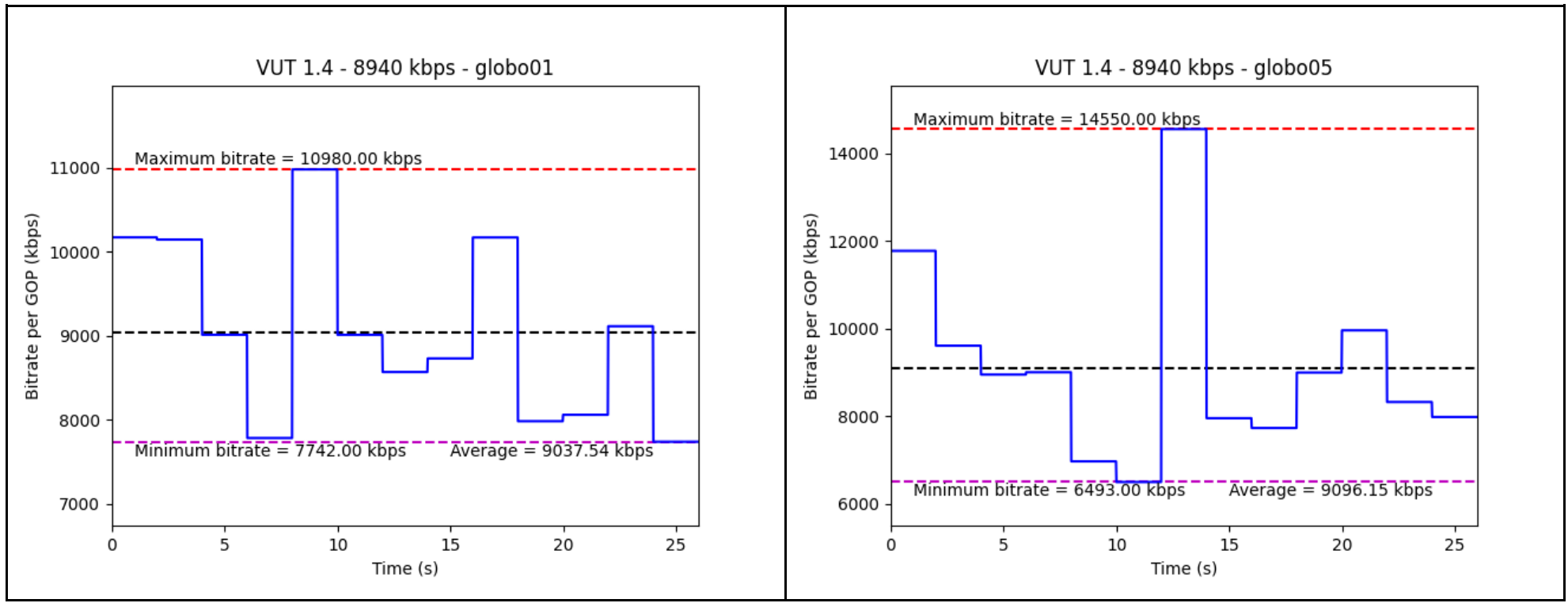

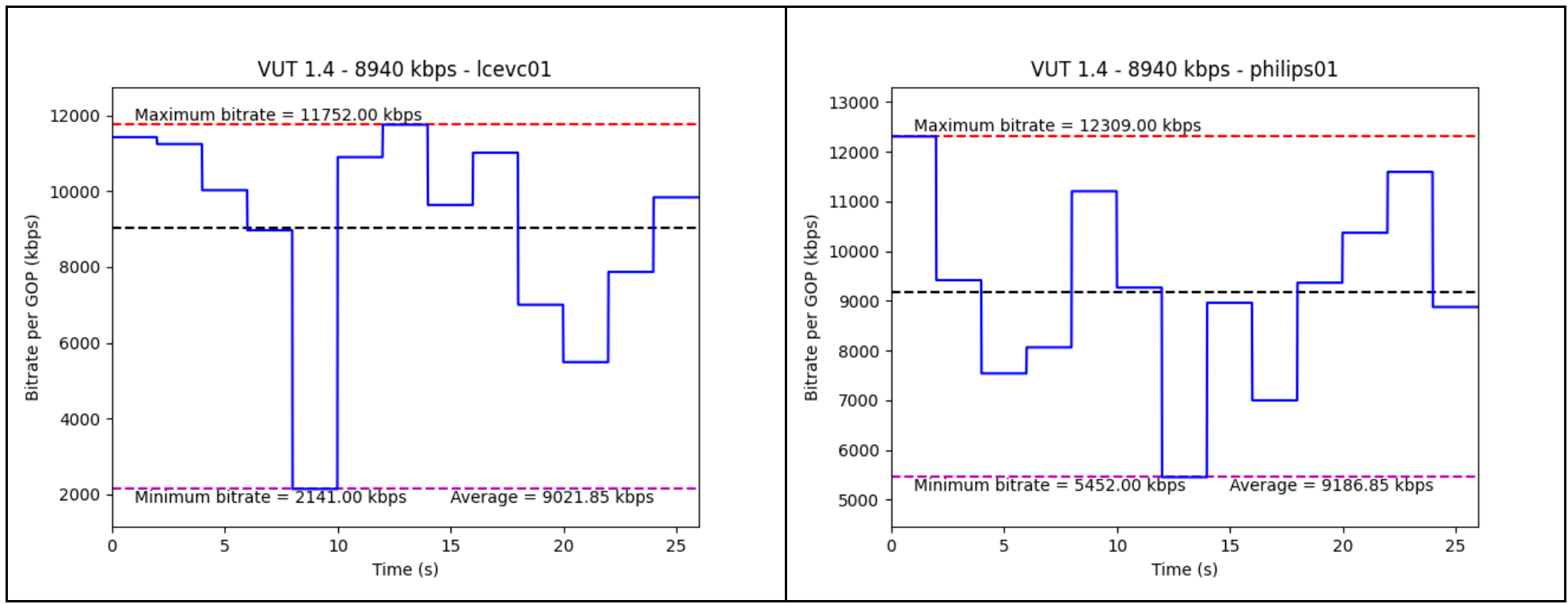

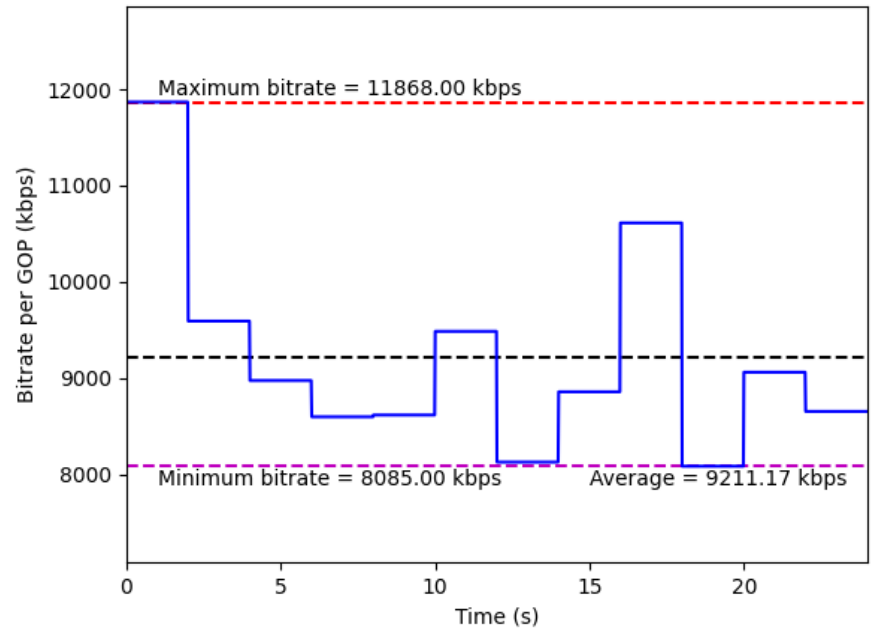

**Bitrate per GOP analysis for target bitrate of 13370 kbps**

*Table 16 - VUT 1.4 Bitrate per GOP when encoded with the output bitrate of 13.37 Mbps*

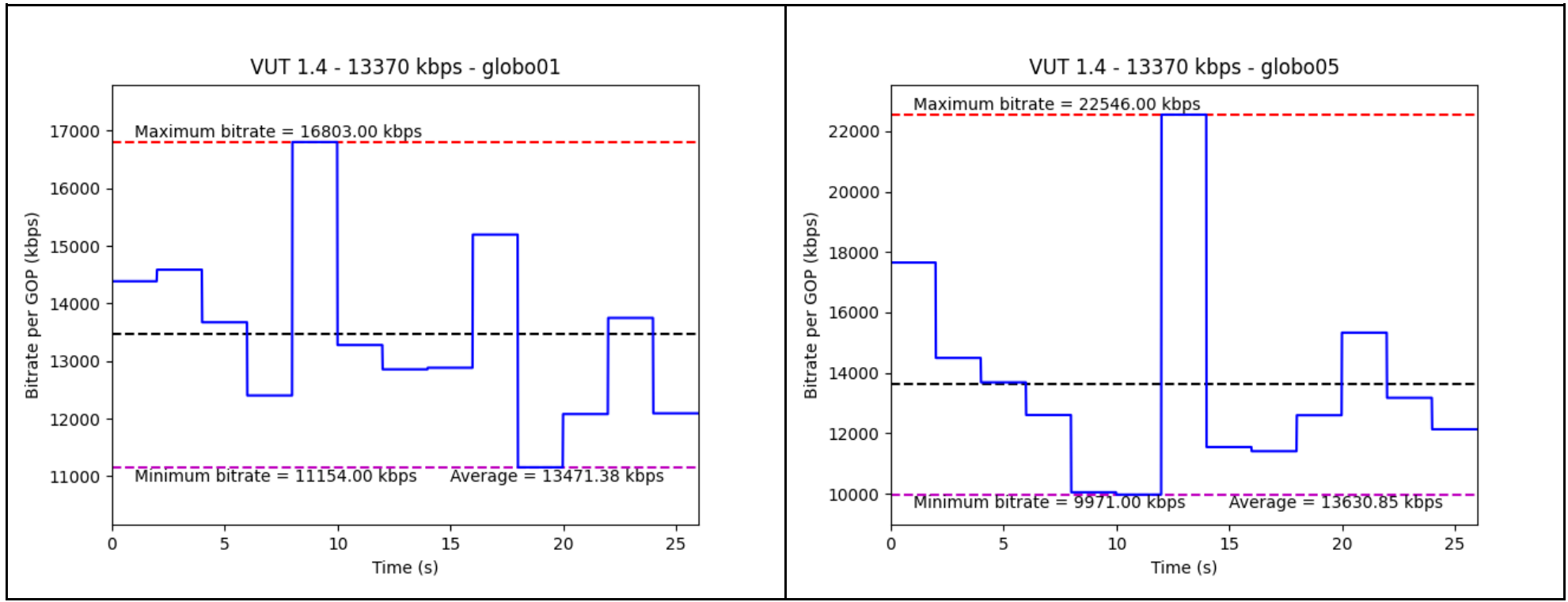

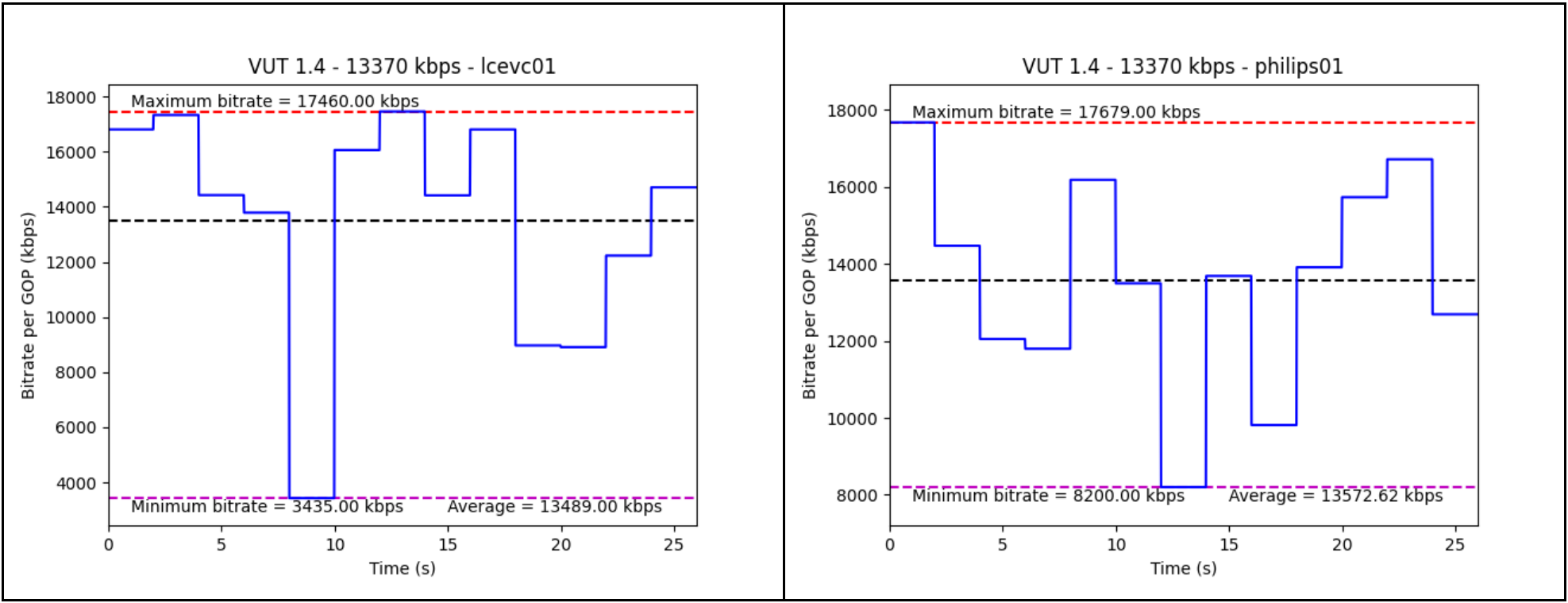

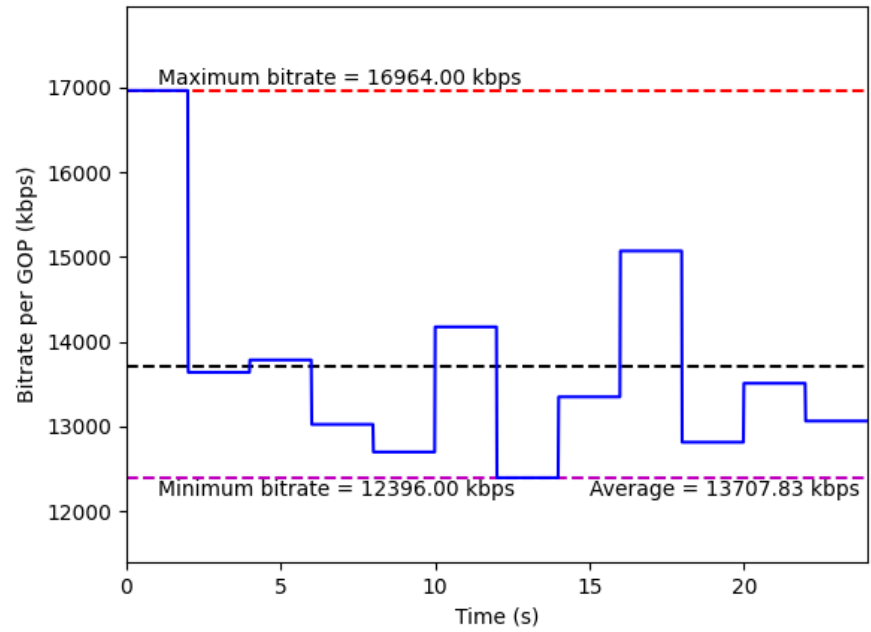

**Bitrate per GOP analysis for target bitrate of 20000 kbps**

*Table 17 - VUT 1.4 Bitrate per GOP when encoded with the output bitrate of 20.00 Mbps*

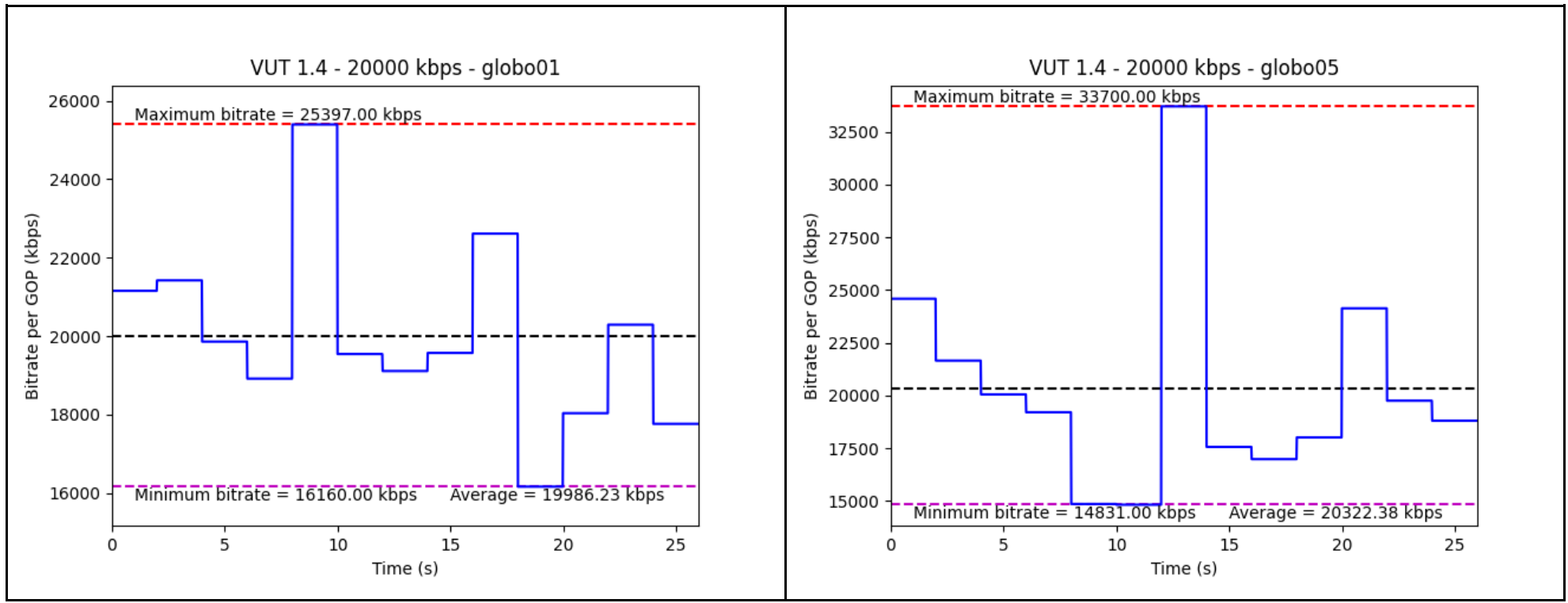

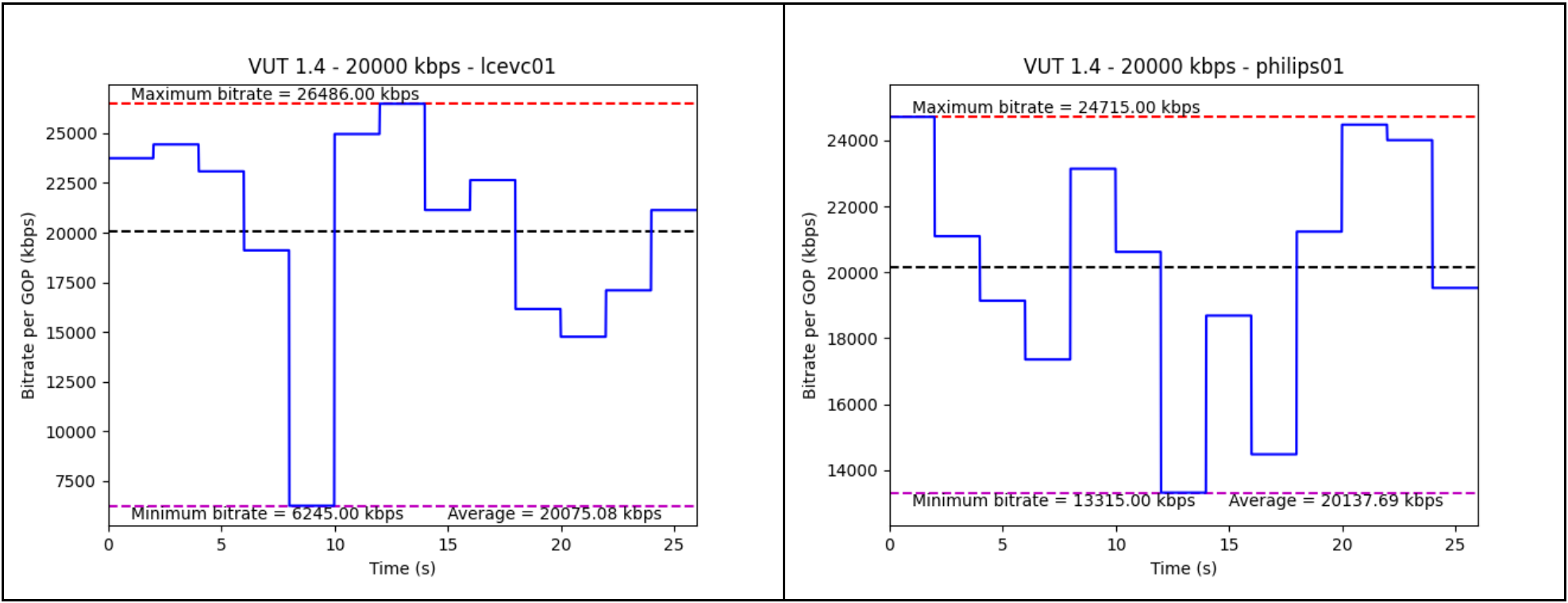

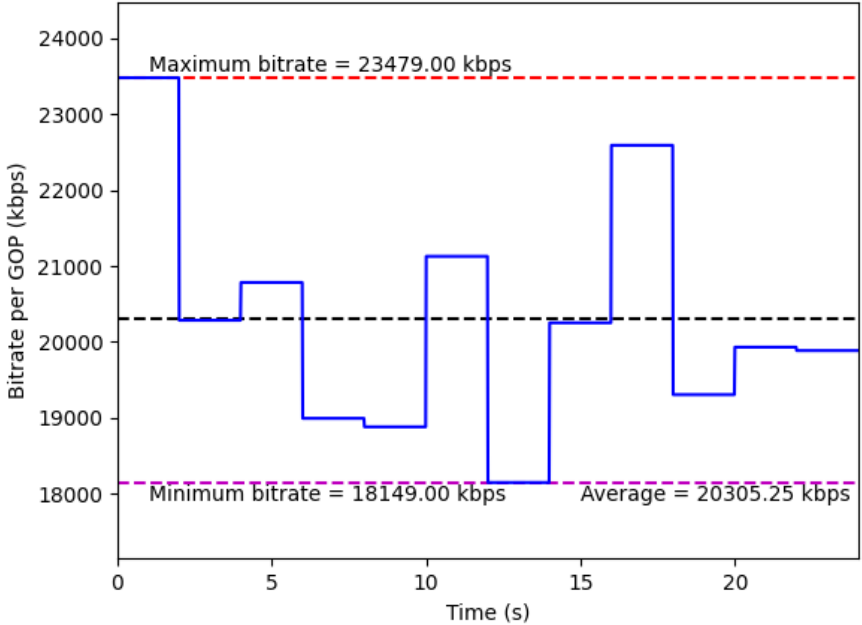

**Bitrate per GOP Summary**

*Table 18 - Average bitrate for all bitstreams tested in VUT 1.4. All bitrates are in kbps.*

| Target Bitrate | Content | Average Rate |
|---|---|---|
| 4000 | globo01 | 4061 |
| | globo05 | 4064 |
| | lcevc01 | 4040 |
| | philips01 | 4161 |
| | philips03 | 4117 |
| 5980 | globo01 | 6054 |
| | globo05 | 6106 |
| | lcevc01 | 6066 |
| | philips01 | 6183 |
| | philips03 | 6164 |
| 8960 | globo01 | 9037 |
| | globo05 | 9096 |
| | lcevc01 | 9021 |
| | philips01 | 9186 |
| | philips03 | 9211 |
| 13333 | globo01 | 13471 |
| | globo05 | 13630 |
| | lcevc01 | 13489 |
| | philips01 | 13572 |
| | philips03 | 13707 |
| 20000 | globo01 | 19986 |
| | globo05 | 20322 |
| | lcevc01 | 20075 |
| | philips01 | 20137 |
| | philips03 | 20305 |

As can be seen from the table, the average bitrate per GOP is very close to the target bitrate.

### 4.3.1.4 VUT 1.4 Analysis and Conclusions

As depicted in Table 12, the output target bitrate for VUT 1.4 for a target quality of -1 ("slightly worse") was found to be 10.90 Mbps. This result is based on the content "globo05" that demanded the highest bitrate. The output target bitrate for a target quality of 0 ("same quality") was achieved at 16.58 Mbps, also for "globo05" video sequence. The output target bitrate for a target quality of 1 ("slightly higher") could not be achieved for "globo05", "lcevc0'", "philips01" and "philips03" contents. As it was done in the 2023 tests, we considered as output target bitrate the highest rate achieved with "same quality" (0). Note that using the highest bitrate corresponding to the target score 0 ("the same") among the five clips in the test material means that this VUT would provide a similar subjective quality for the clip with the highest required bitrate, and somewhat higher score in the other clips while not necessarily reaching the score 1 ("slightly better") in all clips.

### 4.3.2 VUT 2.5 Definition

The goal of this VUT was to test the VVC+LCEVC encoder working at 2 160p resolution. Similar to the 2023 tests for the VUT 2.4, the reference video for this VUT was encoded with the VVC at 1 080p resolution, at 7.52 Mbps. The complete details are found in Table 34. Since the VUT was encoded with a higher resolution than the reference video, the quality target considered was "the same" (0). Note that using the highest bitrate corresponding to the target score 0 ("the same") among the five clips in the test material means that this VUT would provide a similar subjective quality for the clip with the highest required bitrate, and somewhat higher score in the other clips while not necessarily reaching the score 1 ("slightly better") in all clips.

*Table 19 - VUT 2.5 encoding details*

|  | **Reference Video** | **VUT 2.5** |
|---|---|---|
| **Content label** | 1 080p | 2 160p |
| **Resolution** | 1 920 x 1 080 | 3 840 x 2 160 |
| **Frame rate** | 59.94 fps | 59.94 fps |
| **Scan** | Progressive | Progressive |
| **Bit depth** | 10 bits | 10 bits |
| **Color gamut** | BT.2020 | BT.2020 |
| **HDR Mode** | HDR10 | HDR10 |
| **Codec** | H.266/VVC | H.266/VVC + LCEVC |
| **GOP size** | 120 frames (2 seconds) | 120 frames (2 seconds) |
| **Encoder** | Ateme TitanLive Innovation v 4.1.31.911 | MainConcept Live Encoder v 0.0.0.4036 |
| **Encoder type** | Real-time | Real-time |
| **Bitrate** | 7.52 Mbps | [2.67 3.99 5.96 8.92 13.33] Mbps[2] |

---

[2] The bitrates consider the total bitrate for the VVC base layer and LCEVC enhancement layer. The decoded video also decodes both layers, outputting a 4K resolution video.

### 4.3.2.1 VUT 2.5 Experimental Findings

**Threshold σ=-1**

*Table 20 - VUT 2.5 results targeting a MOS grade of -1 (slightly worse)*

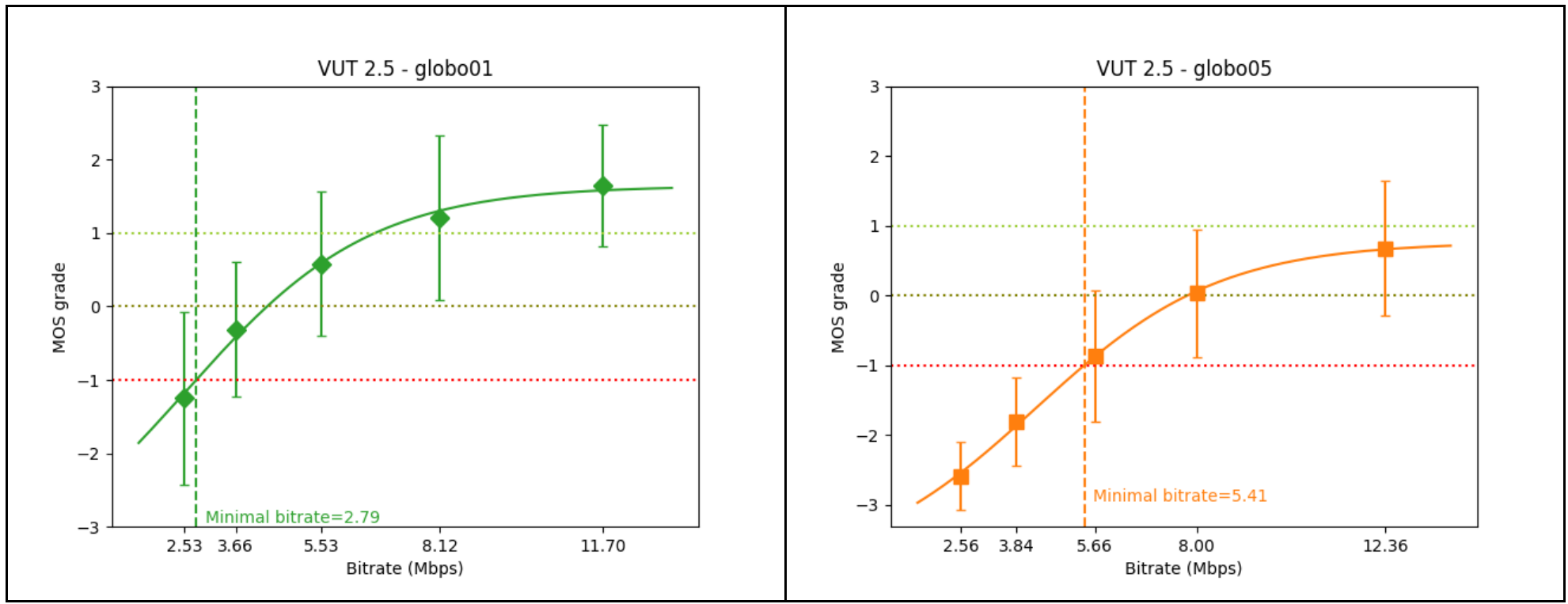

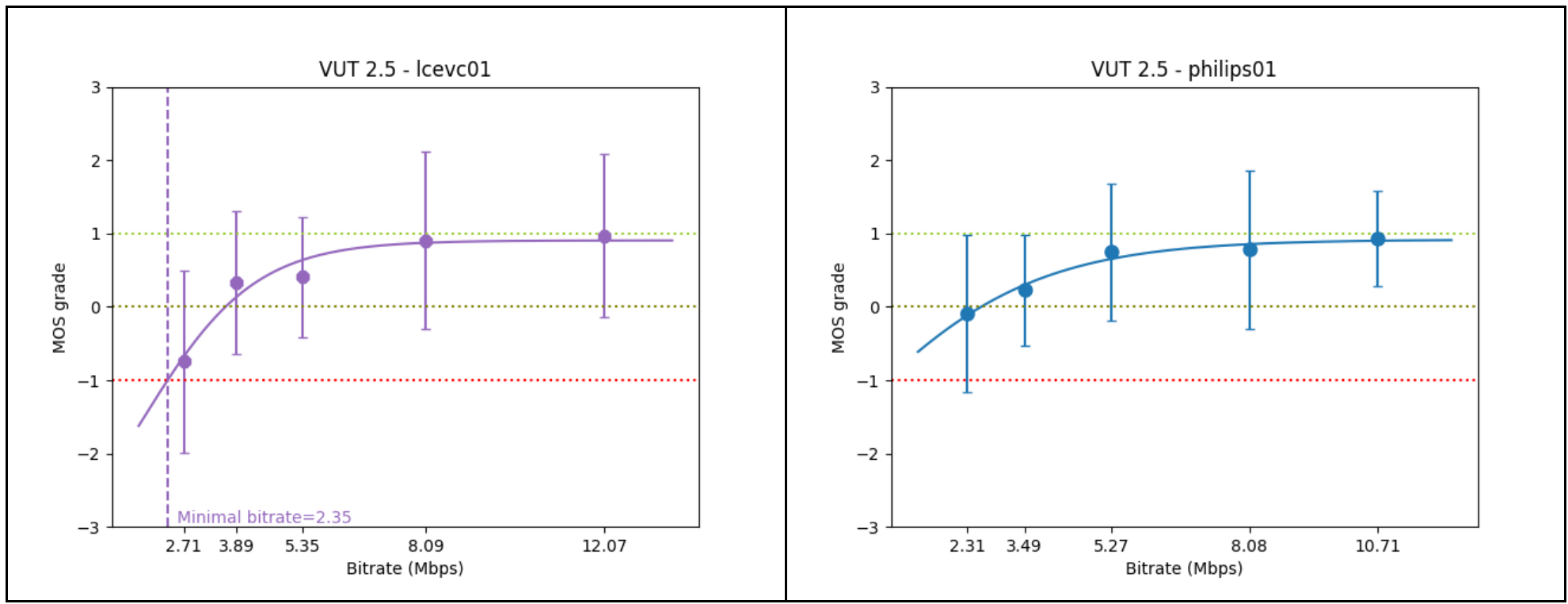

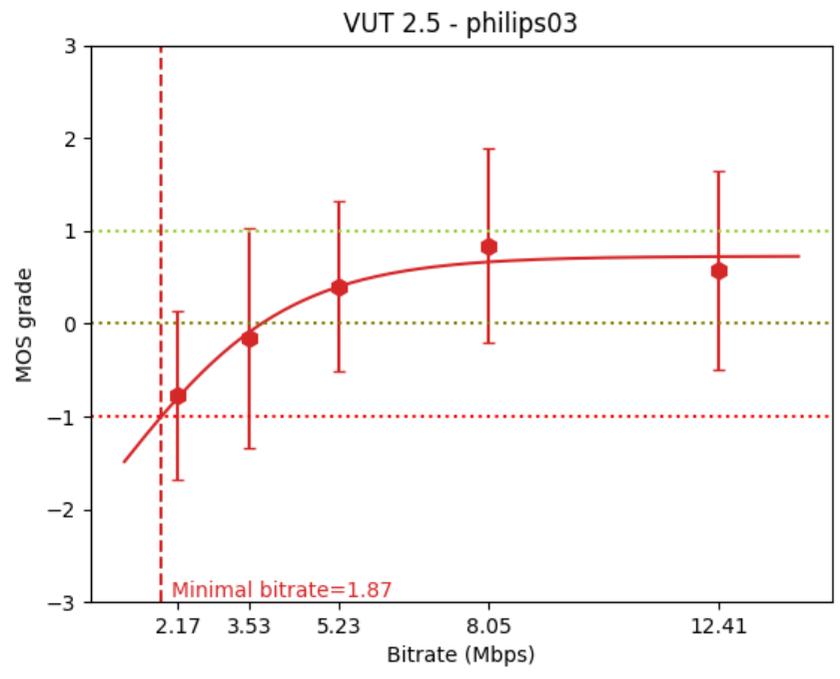

**Threshold σ=0**

*Table 21 - VUT 2.5 results targeting a MOS grade of 0 (same quality)*

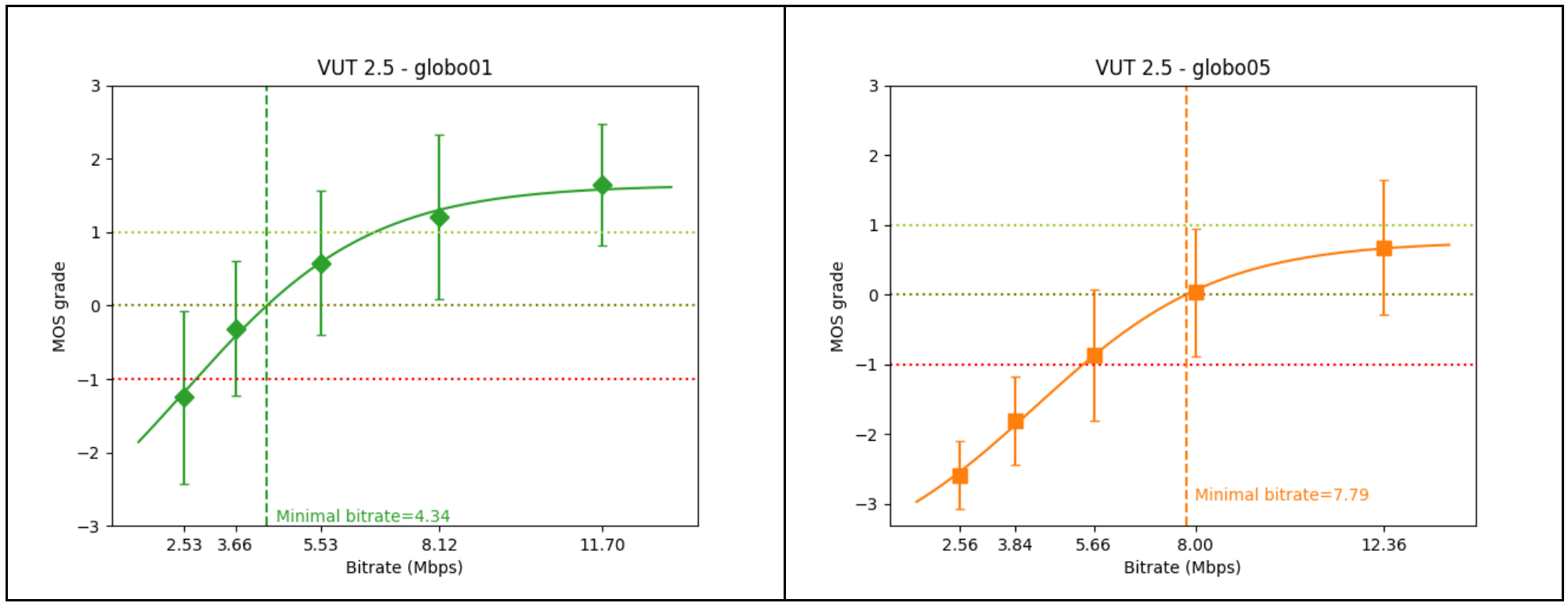

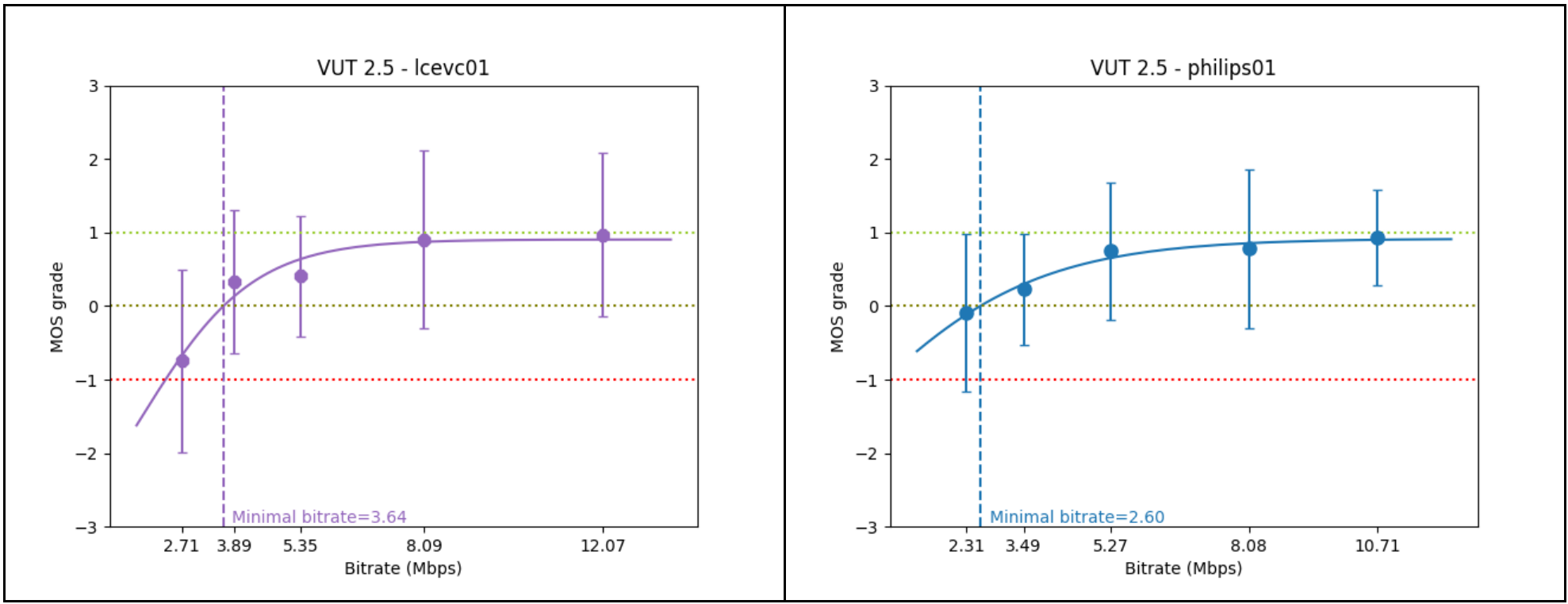

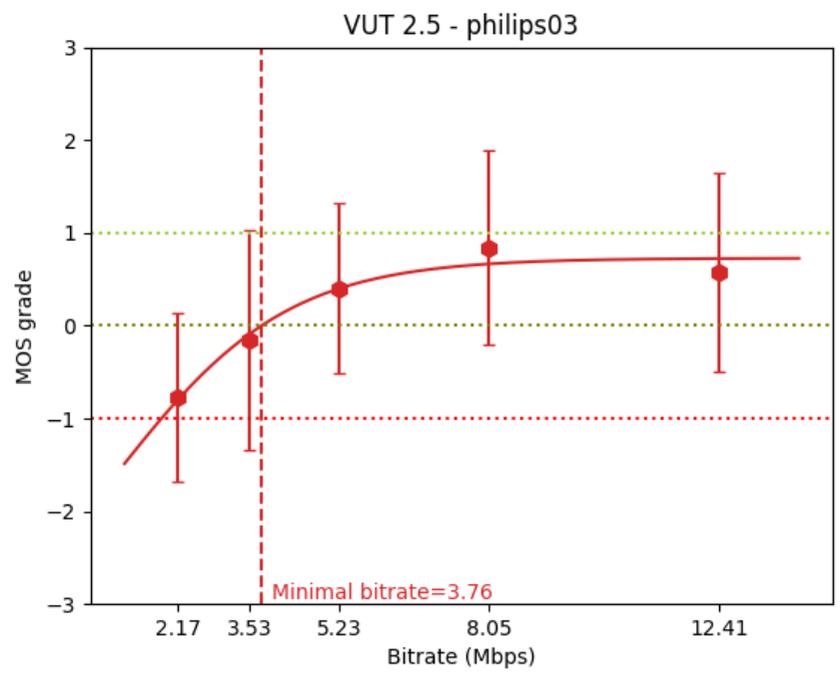

**Threshold σ=1**

Table 22 - VUT 2.5 results targeting a MOS grade of 1 (slightly better)

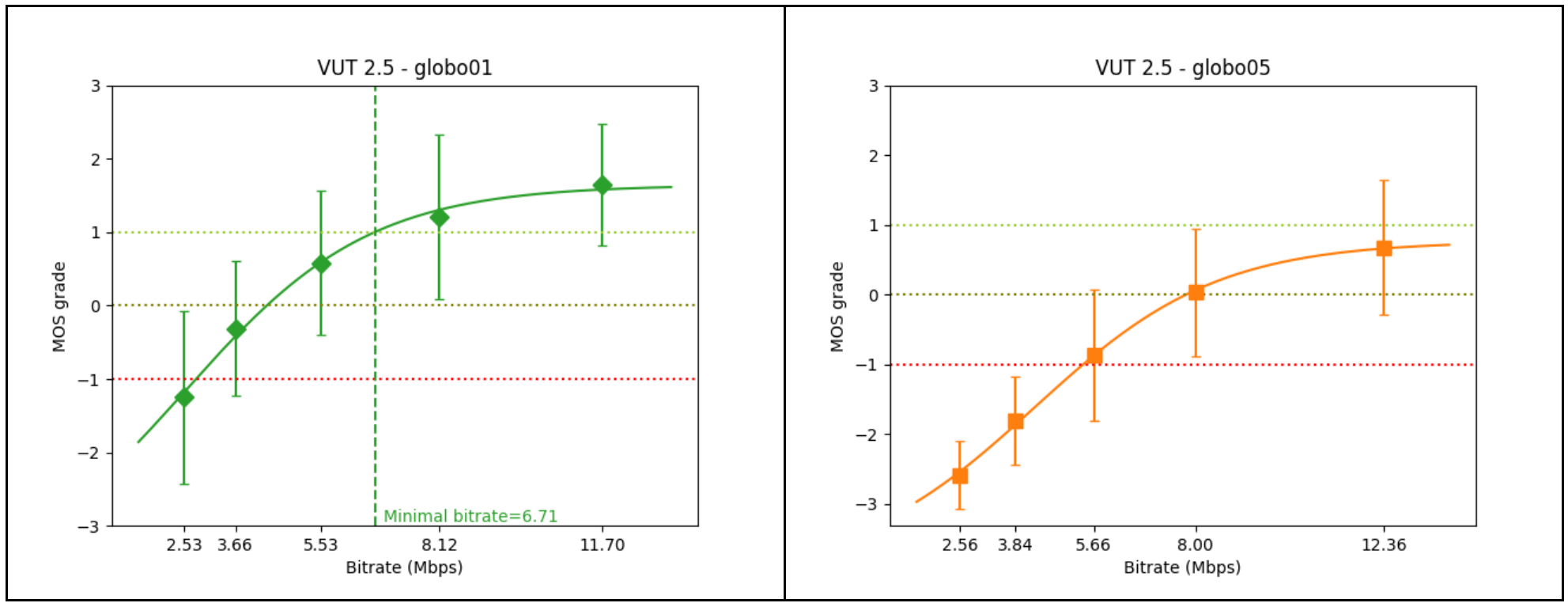

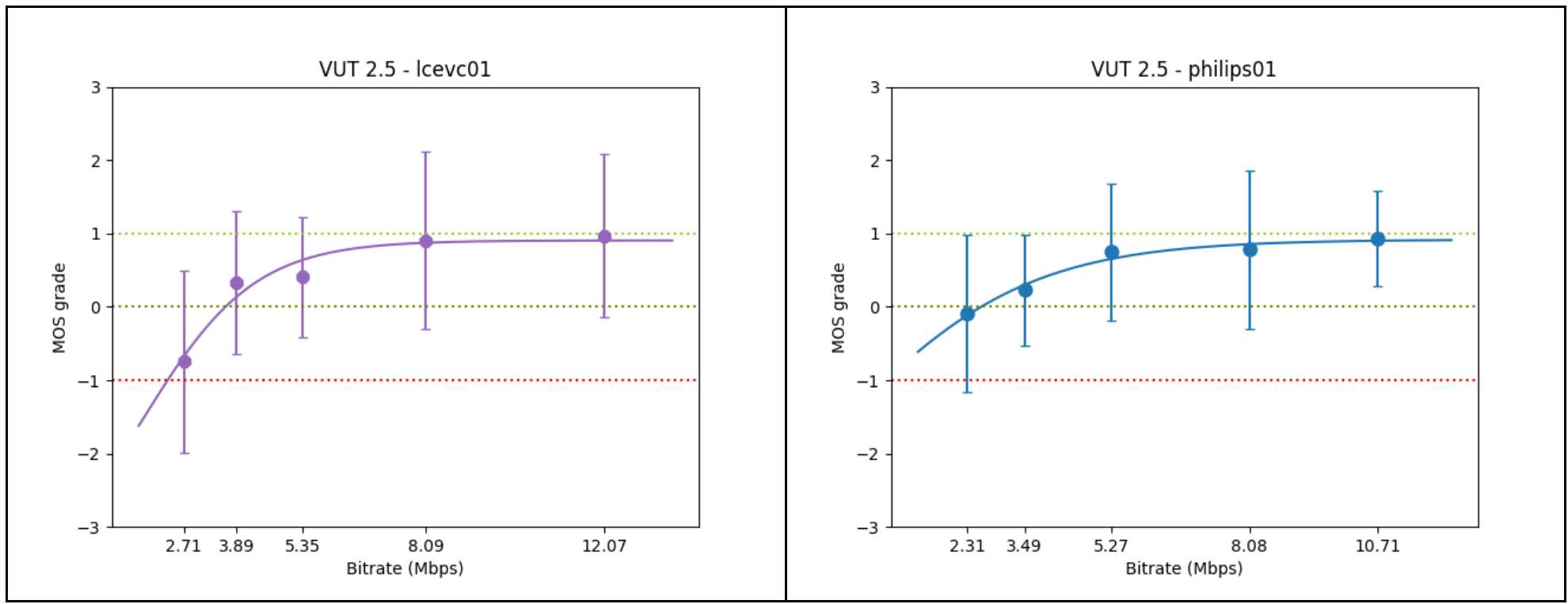

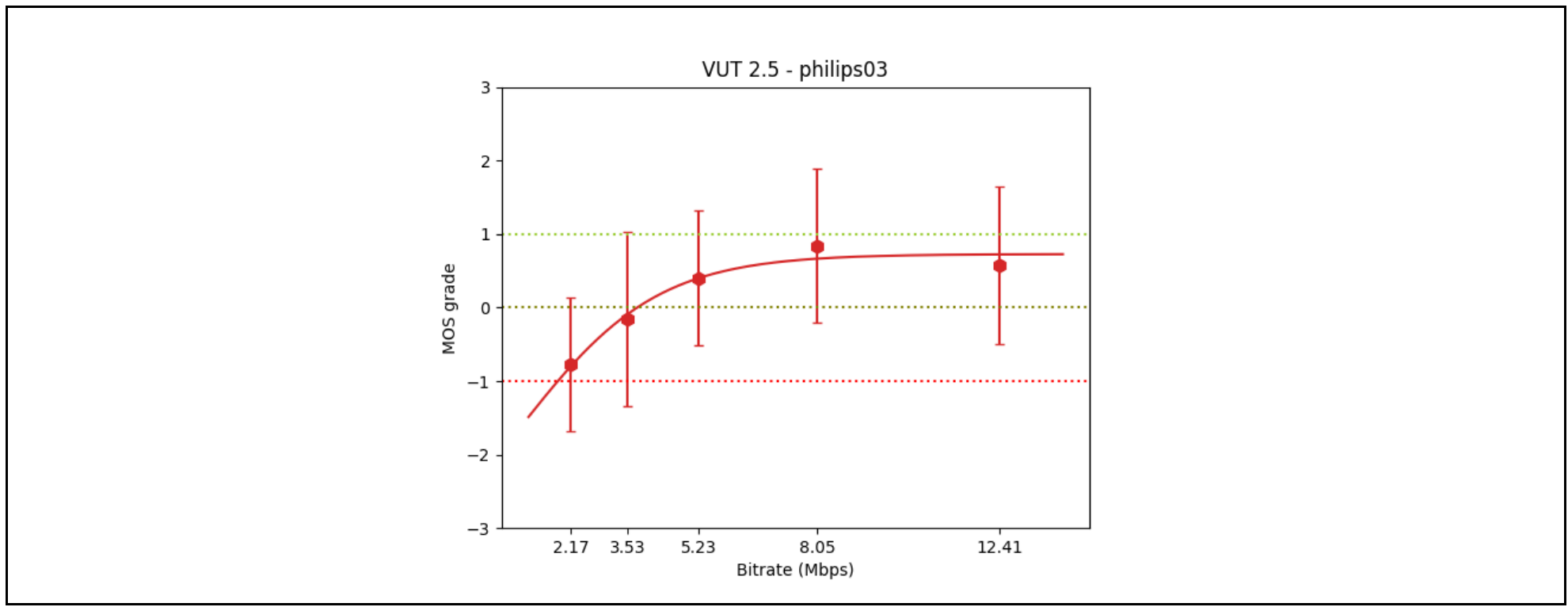

### 4.3.2.2 VUT 2.5 Output bitrate analysis

The bitrates for which each target quality is achieved are shown in Table 23.

*Table 23 - Achieved Target Bitrates per content (the output target bitrate is highlighted)*

| Target Quality | globo01 | globo05 | Icevc01 | philips01 | philips03 |
|---|---|---|---|---|---|
| -1 | 2.79 Mbps | 5.41 Mbps | 2.35 Mbps | Achieved for all bitrates | 1.87 Mbps |
| 0 | 4.34 Mbps | **7.79 Mbps** | 3.64 Mbps | 2.60 Mbps | 3.76 Mbps |
| 1 | 6.71 Mbps | Not achieved | Not achieved | Not achieved | Not achieved |

Following the results from the previous sections, the output bitrate for VUT 2.5 was found to be 7.79 Mbps.

### 4.3.2.3 VUT 2.5 Bitrate per GOP analysis

We have also carried out an analysis of the bitrate per GOP for each of the bitstreams used in the tests. This is a simple analysis, where the bitrate used for each GOP (i.e., between two intra frames) is computed. The total rate (VVC base layer and LCEVC enhancement layer) is considered.

**Bitrate per GOP analysis for target bitrate of 2667 kbps**

*Table 24 - VUT 2.5 Bitrate per GOP when encoded with the output bitrate of 2.67 Mbps*

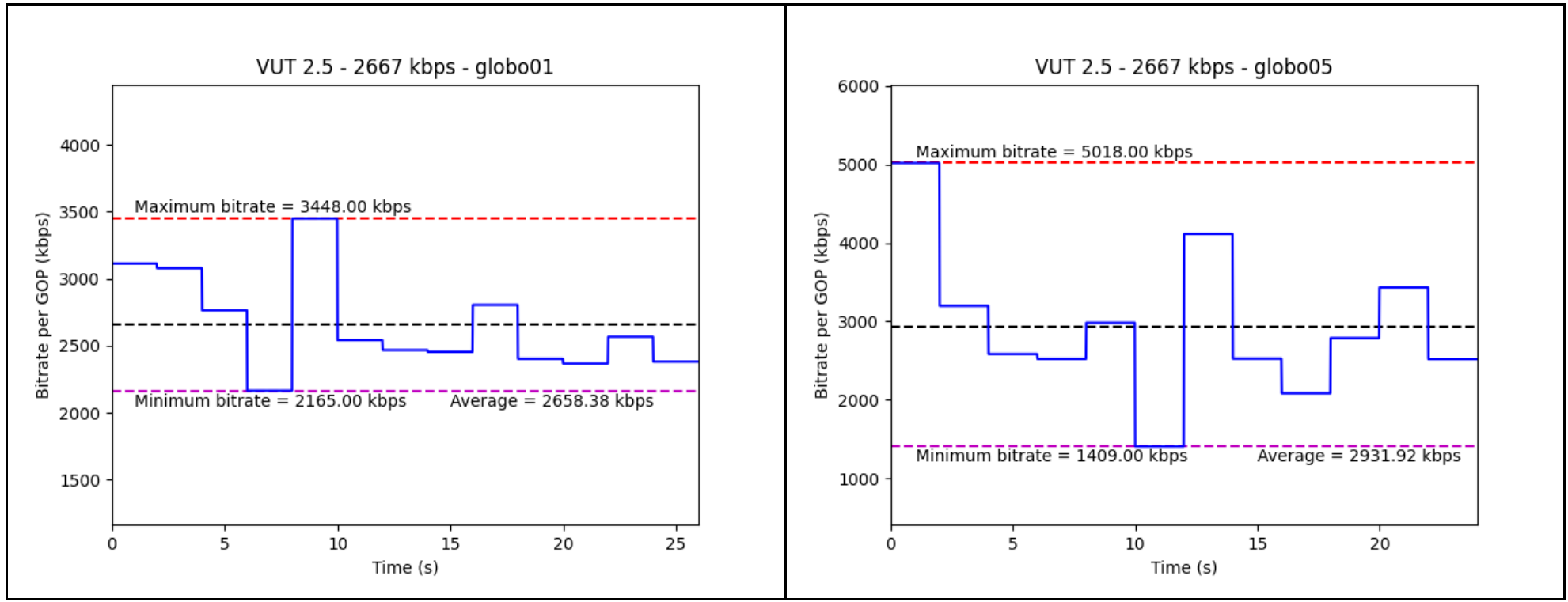

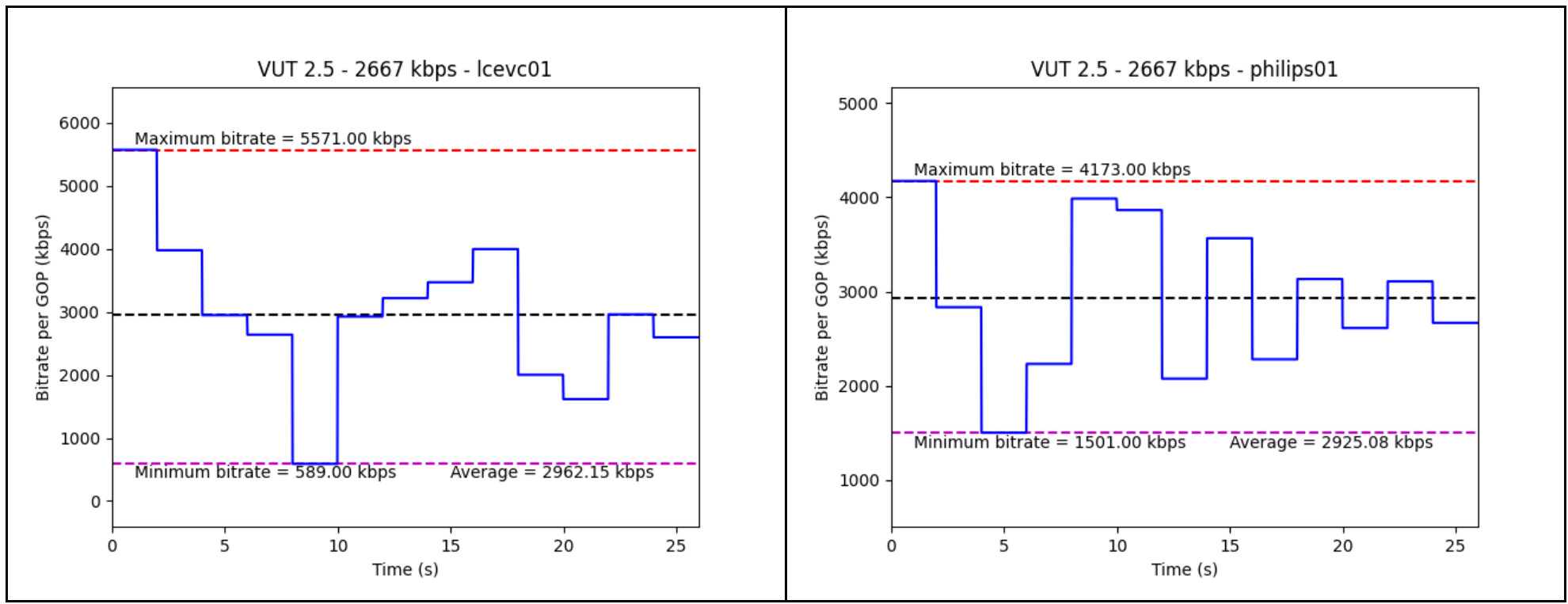

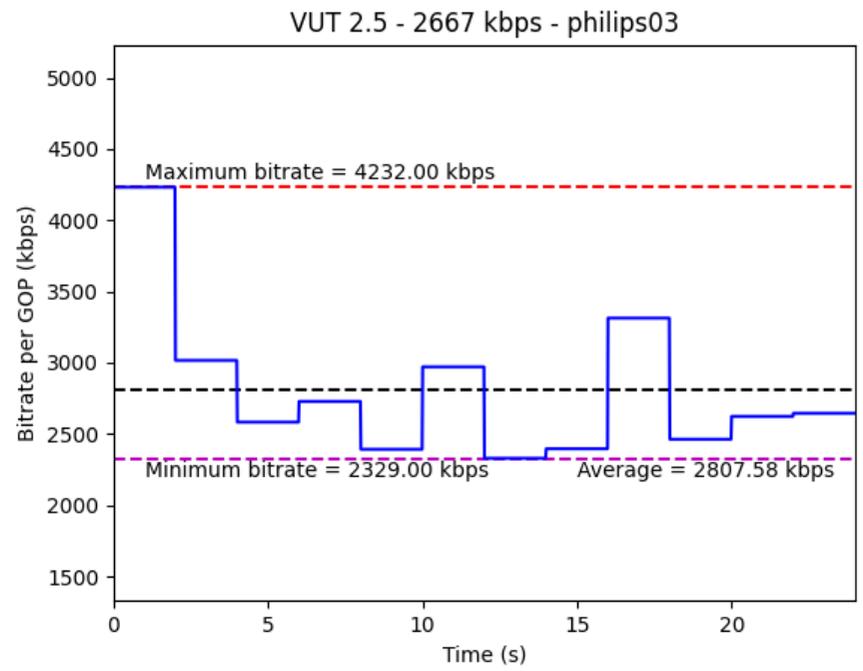

**Bitrate per GOP analysis for target bitrate of 3987 kbps**

*Table 25 - VUT 2.5 Bitrate per GOP when encoded with the output bitrate of 3.98 Mbps*

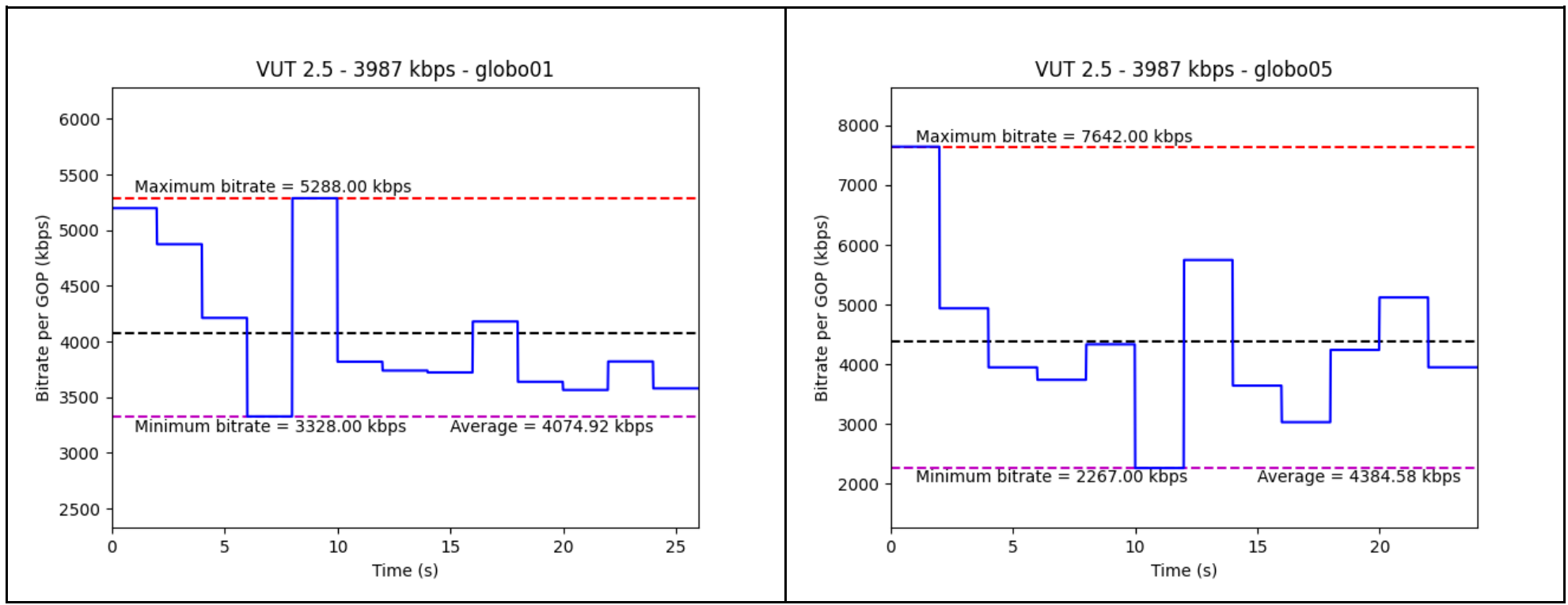

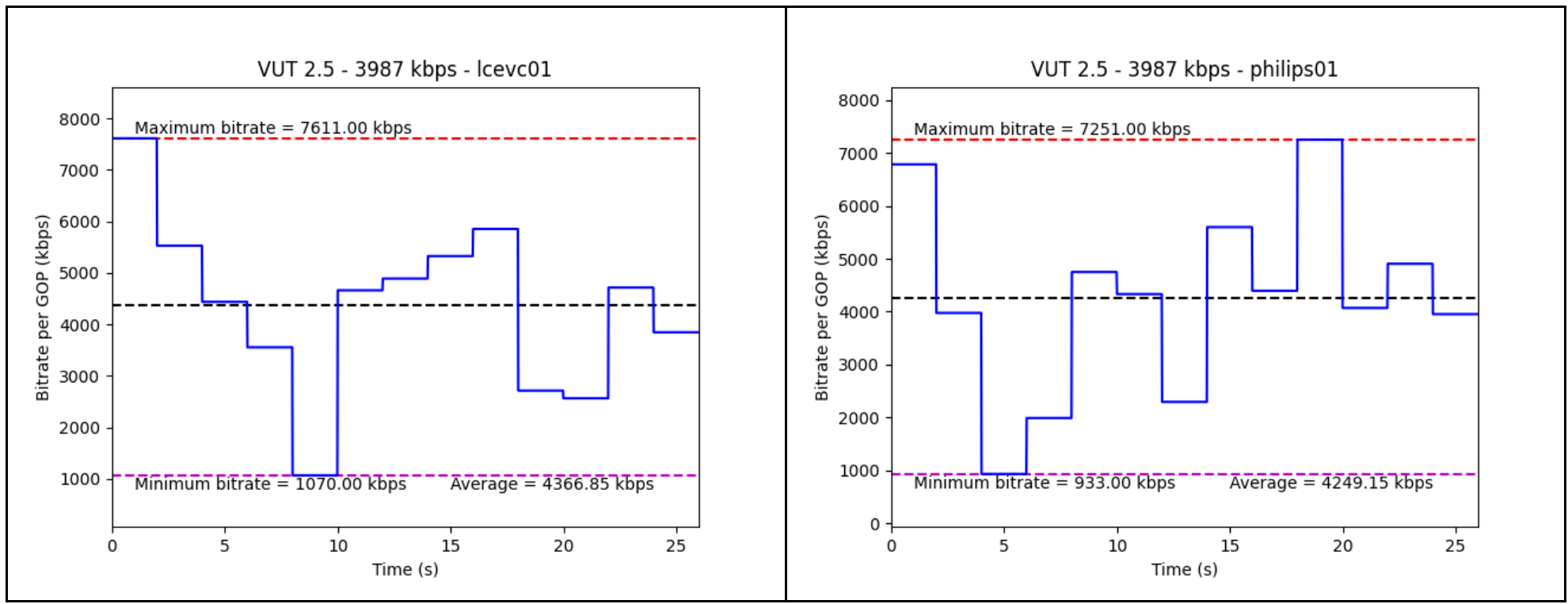

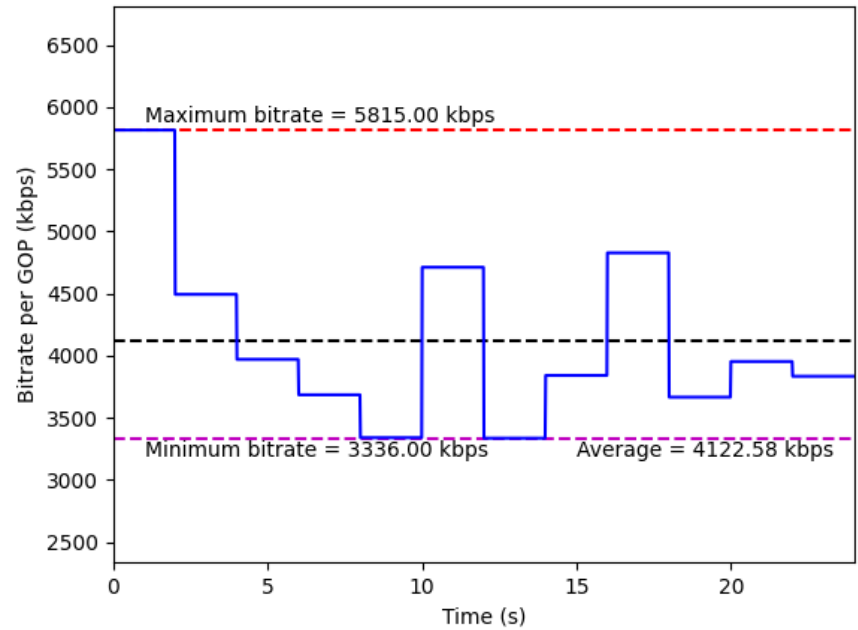

**Bitrate per GOP analysis for target bitrate of 5960 kbps**

*Table 26 - VUT 2.5 Bitrate per GOP when encoded with the output bitrate of 5.96 Mbps*

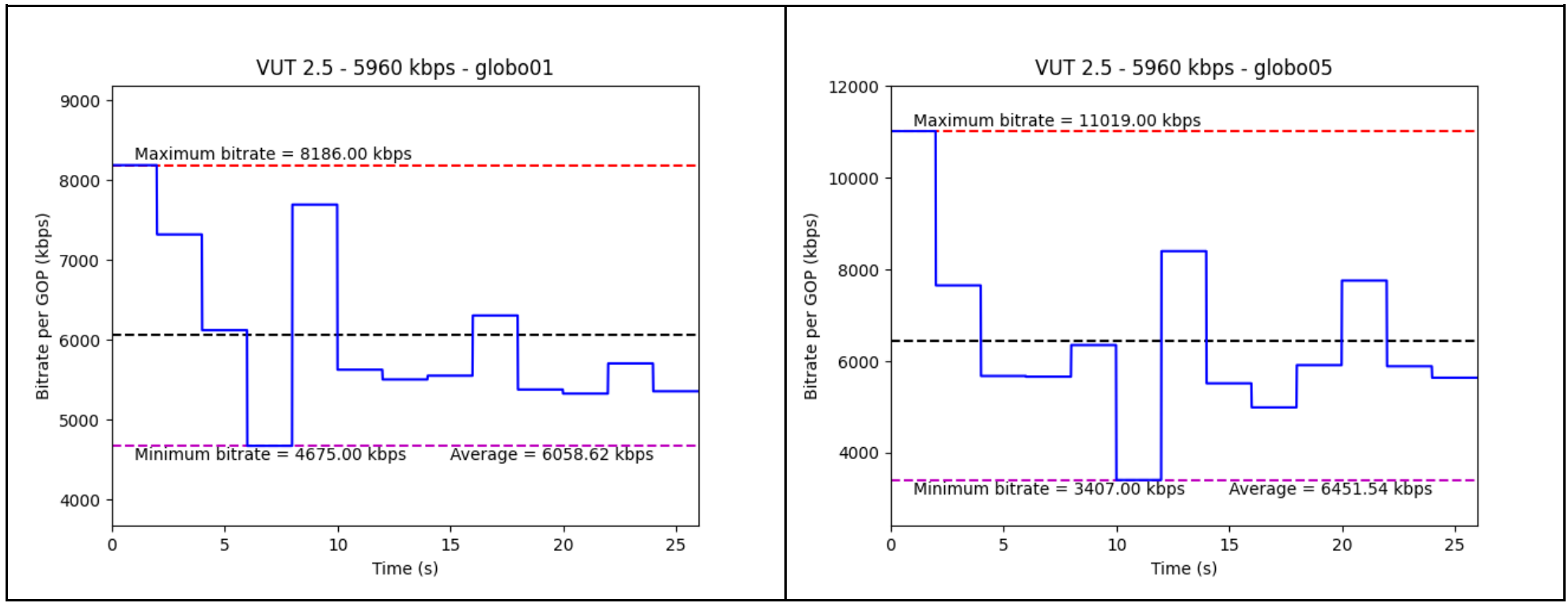

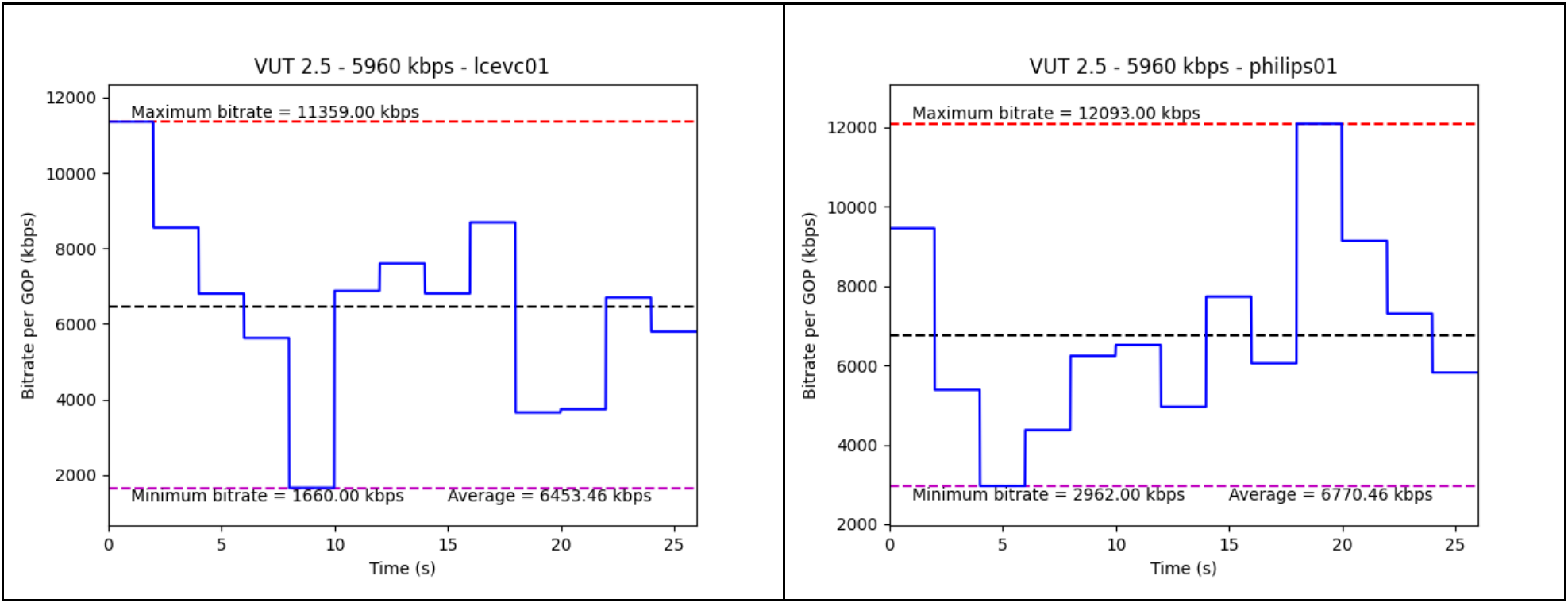

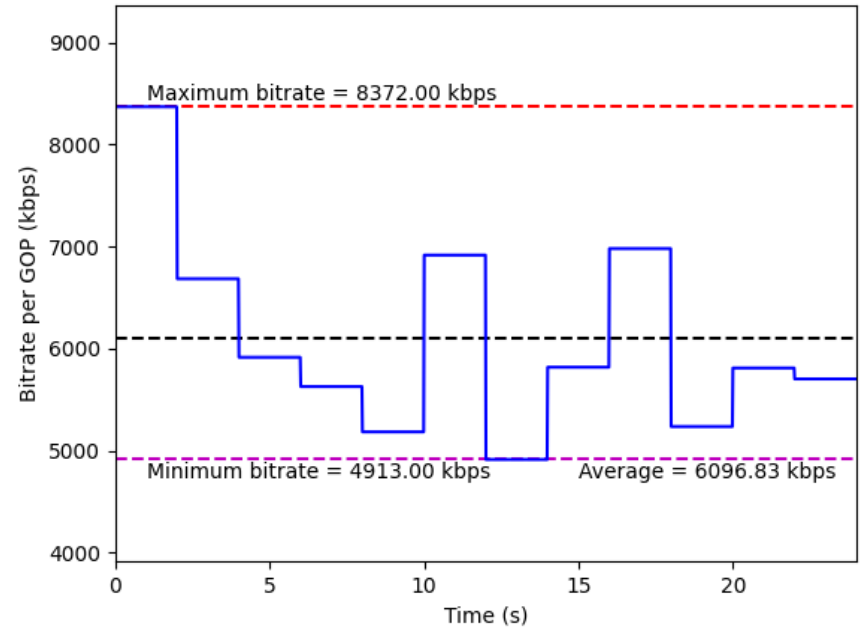

**Bitrate per GOP analysis for target bitrate of 8920 kbps**

*Table 27 - VUT 2.5 Bitrate per GOP when encoded with the output bitrate of 8.92 Mbps*

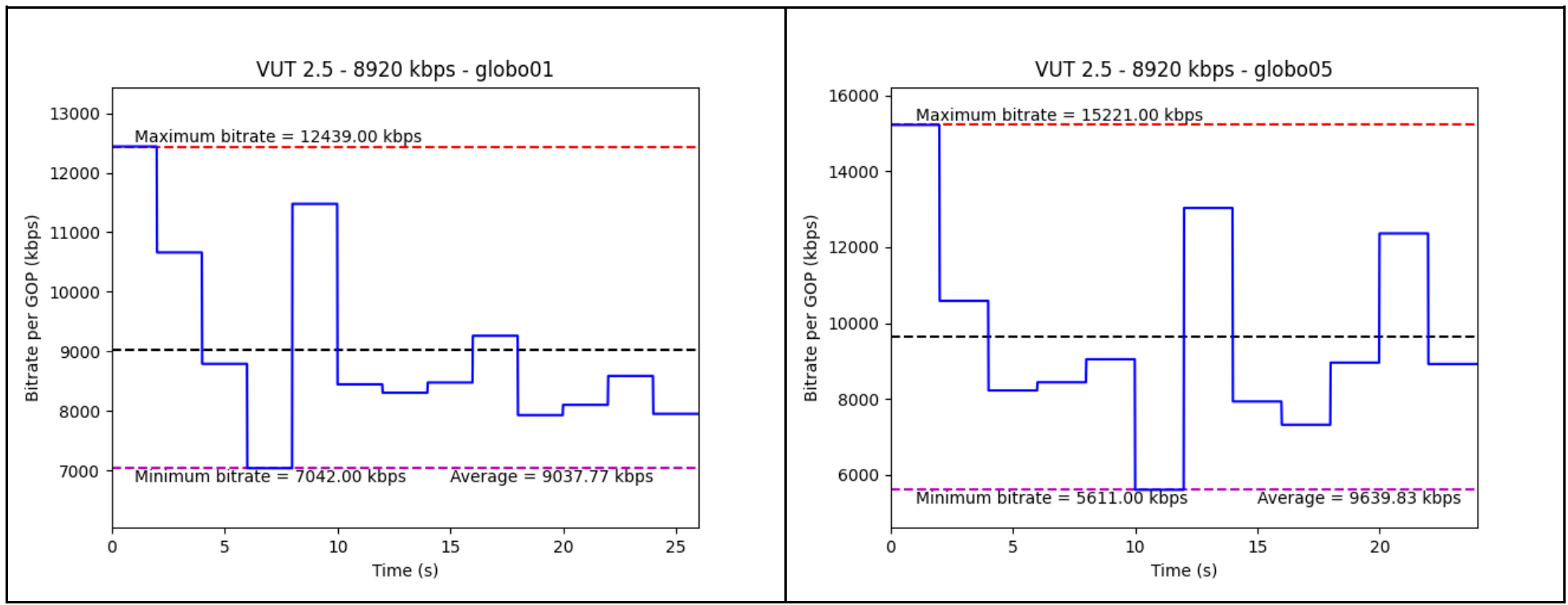

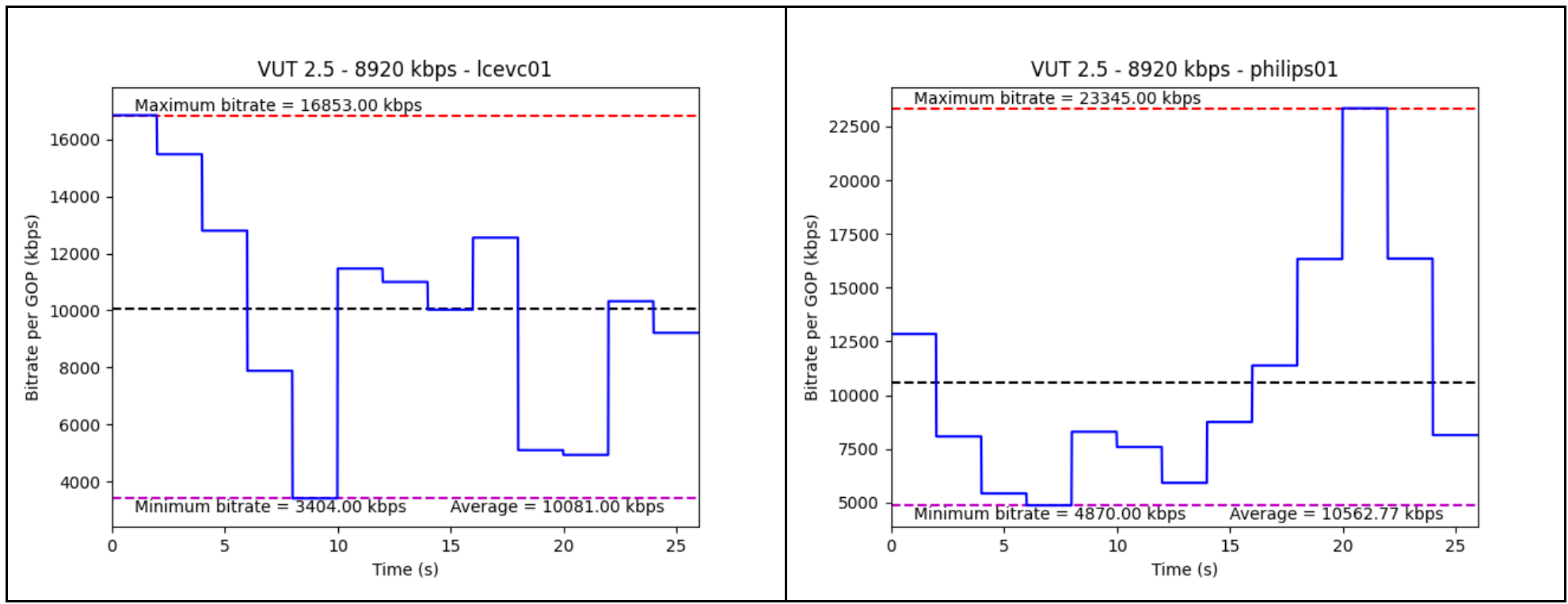

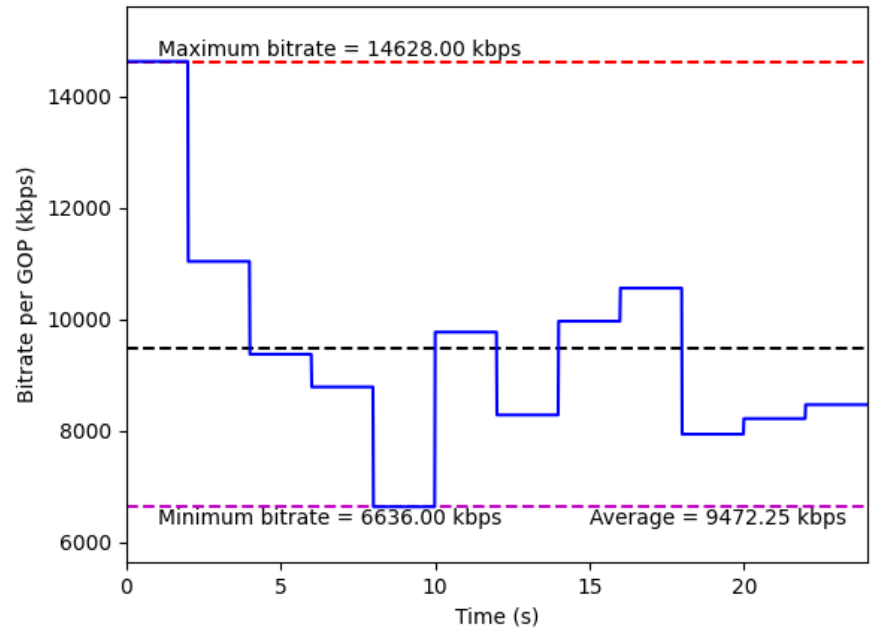

**Bitrate per GOP analysis for target bitrate of 13333 kbps**

Table 28 - VUT 2.5 Bitrate per GOP when encoded with the output bitrate of 13.33 Mbps

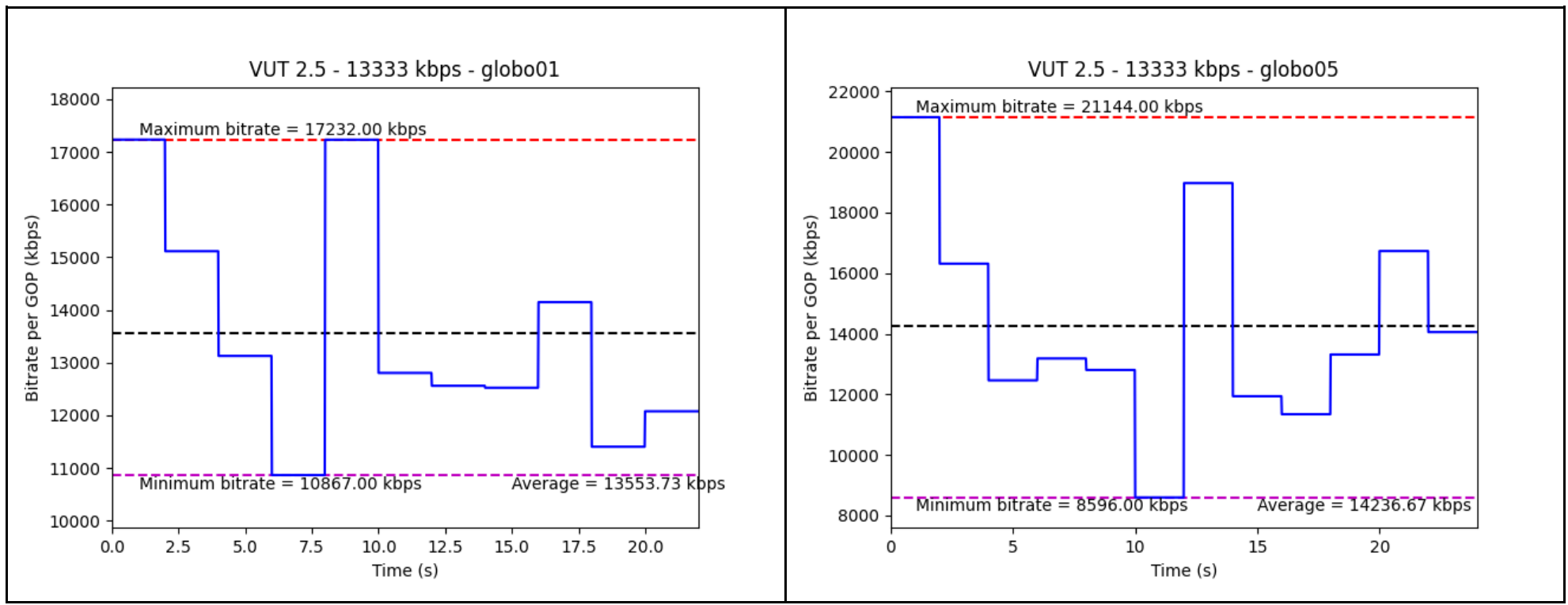

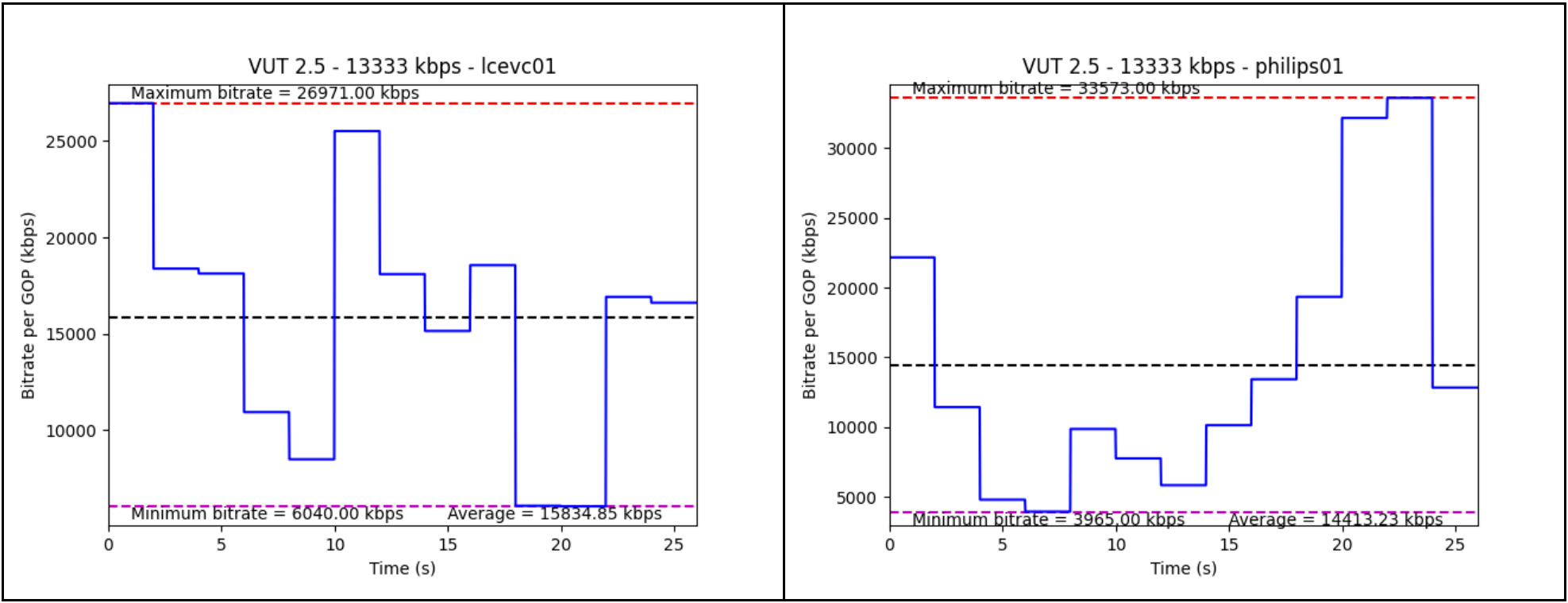

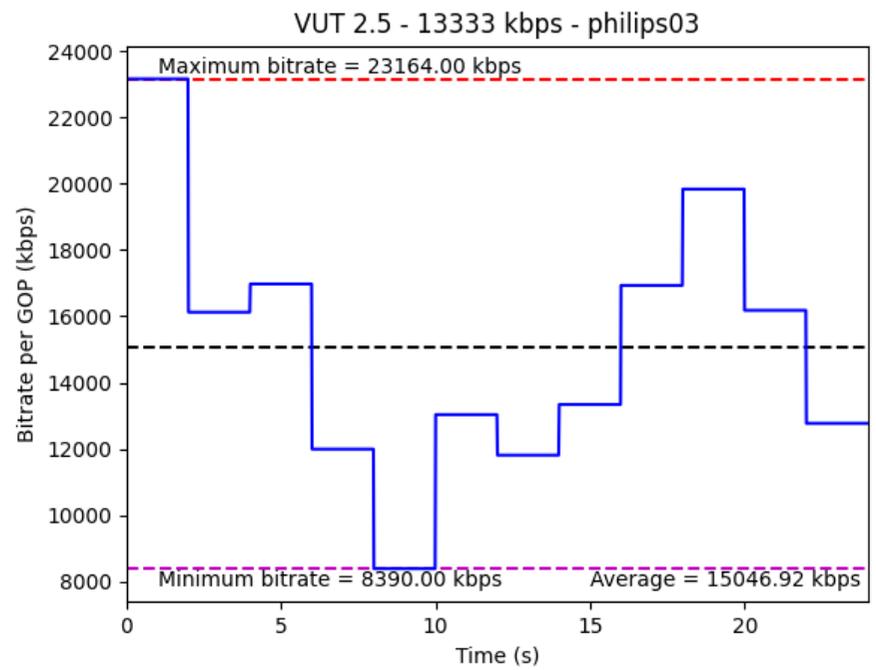

**Bitrate per GOP Summary**

*Table 29 - Average bitrate for all bitstreams tested in VUT 2.5. All bitrates are in kbps.*

| Target Bitrate | Content | Average Rate |
|---|---|---|
| 2667 | globo01 | 2658 |
| | globo05 | 2931 |
| | lcevc01 | 2962 |
| | philips01 | 2925 |
| | philips03 | 2807 |
| 3987 | globo01 | 4074 |
| | globo05 | 4384 |
| | lcevc01 | 4366 |
| | philips01 | 4249 |
| | philips03 | 4122 |
| 5960 | globo01 | 6058 |
| | globo05 | 6451 |
| | lcevc01 | 6453 |
| | philips01 | 6770 |
| | philips03 | 6096 |
| 8920 | globo01 | 9037 |
| | globo05 | 9639 |
| | lcevc01 | 10081 |
| | philips01 | 10562 |
| | philips03 | 9472 |
| 13333 | globo01 | 13553 |
| | globo05 | 14236 |
| | lcevc01 | 15834 |
| | philips01 | 14413 |
| | philips03 | 15046 |

As can be seen from the table, the average bitrate per GOP is close to the target bitrate.

### 4.3.2.4 VUT 2.5 Analysis and Conclusions

As depicted in Table 23, the output target bitrate for VUT 2.5 for a target quality of -1 ("slightly worse") was found to be 5.41 Mbps. This result is based on the content "globo05" that demanded the highest bitrate. The output target bitrate for a target quality of 0 ("same quality") was achieved at 7.79 Mbps, also for "globo05" video sequence. The output target bitrate for a target quality of 1 ("slightly higher") could not be achieved for "globo05", "lcevc0'", "philips01" and "philips03" contents. As it was done in the 2023 tests, we considered as output target bitrate the highest rate achieved with "same quality" (0). Note that using the highest bitrate corresponding to the target score 0 ("the same") among the five clips in the test material means that this VUT would provide a similar subjective quality for the clip with the highest required bitrate, and somewhat higher score in the other clips while not necessarily reaching the score 1 ("slightly better") in all clips.

## 4.4 Summary of the Results

The results of the previous sections are summarized in Table 30. The main output for each VUT is the minimum bitrate for which that encoding configuration achieves that particular quality for all 5 contents tested. Not all VUTs are able to achieve some of the target grades due to the resolution of the sequences tested and the capabilities of the encoders used. The target grade intended is highlighted in the table.

It is important to note that in these tests both VVC encoders use the same real-time software encoder, and both VUTs follow the same configuration.

*Table 30 - Summary of the results*

| VUT | Encoder | Reference Video | Target Grade -1 | Target Grade 0 | Target Grade 1 |
|---|---|---|---|---|---|
| VUT 1.4 | VVC 2 160p | 1 080p VVC + HDR10 at 7.52 Mbps | 10.90 Mbps | **16.58 Mbps** | Not achieved |
| VUT 2.5 | VVC + LCEVC 2 160p | 1 080p VVC + HDR10 at 7.52 Mbps | 5.41 Mbps | **7.79 Mbps** | Not achieved |

These bitrate values were determined considering worst-case situations. Note that using the highest bitrate corresponding to the target score 0 ("the same") among the five clips in the test material means that a given VUT would provide a similar subjective quality for the clip with the highest required bitrate, and somewhat higher score in the other clips while not necessarily reaching the score 1 ("slightly better") in all clips.